\documentclass[twocolumn,superscriptaddress,prb,10pt]{revtex4-1}
\usepackage{verbatim}
\usepackage{braket}
\usepackage{amsmath,amssymb}
\usepackage{tikz}
\usepackage{graphicx}
\usepackage{color}
\usepackage[colorlinks,bookmarks=false,citecolor=blue,linkcolor=red,urlcolor=blue]{hyperref}
\usepackage{times}

\begin{document}

\title{Towards a Generalized Hydrodynamics description of R\'enyi entropies in integrable systems}

\author{Vincenzo Alba}
\affiliation{International School for Advanced Studies (SISSA),
Via Bonomea 265, 34136, Trieste, Italy, 
INFN, Sezione di Trieste}

\date{\today}

\begin{abstract} 

We investigate the steady-state R\'enyi entanglement entropies after a 
quench from a piecewise homogeneous initial state in integrable models. 
In the quench protocol two macroscopically different chains 
(leads) are joined together at the initial time, and the subsequent dynamics 
is studied. We study the entropies of a finite subsystem 
at the interface between the two leads. The density of R\'enyi entropies 
coincides with that of the entropies of the Generalized Gibbs 
Ensemble (GGE) that describes the interface between the chains. 
By combining the Generalized Hydrodynamics (GHD) treatment of the 
quench with the Bethe ansatz approach 
for the R\'enyi entropies, we provide exact results for quenches from 
several initial states in the anisotropic Heisenberg chain (XXZ chain), 
although the approach is applicable, in principle, to any low-entangled 
initial state and any integrable model. 
An interesting protocol that we consider is the 
{\it expansion quench}, in which one of the two leads is prepared 
in the vacuum of the model excitations. An intriguing feature is 
that for moderately large anisotropy the transport of bound-state is not 
allowed. Moreover, we show that there is a ``critical'' 
anisotropy, below which bound-state transport 
is permitted. This is reflected in the steady-state entropies, which 
for large enough anisotropy do not contain information about the bound 
states. Finally, we benchmark our results against 
time-dependent Density Matrix Renormalization Group (tDMRG) 
simulations.

\end{abstract}


\maketitle

\section{Introduction} 
\label{sec-intro}

In recent years, entanglement-related quantities emerged as informative 
witnesses of the complexity of the quantum many-body wavefunction, 
both at equilibrium and out of equilibrium. 
Entanglement is deeply intertwined with the computational cost 
to accurately describe the quantum many-body wavefunctin by 
using Matrix Product States~\cite{swvc-08,pv-08,rev0,d-17}. 
The R\'enyi entropies $S^{(\alpha)}$ are popular entanglement 
measures for pure states. Given a subsystem $A'$, they are defined as 
\begin{equation}
\label{renyi-def}
S^{(\alpha)}\equiv \frac{1}{1-\alpha}\ln\textrm{Tr}\rho_{A'}^\alpha, 
\quad\textrm{with}\,\,\alpha\in\mathbb{R}. 
\end{equation}
where $\rho_{A'}$ is the reduced density matrix of $A'$. 
The limit $\alpha\to 1$ in~\eqref{renyi-def} defines the von Neumann 
entropy as $S\equiv -\textrm{Tr}\rho_{A'}\ln\rho_{A'}$. 
The knowledge of R\'enyi entropies for different values of 
$\alpha$ gives access to the distribution of the full spectrum of 
$\rho_{A'}$~\cite{calabrese-2008} (entanglement spectrum). 
Importantly, the R\'enyi entropies can be computed by using quantum 
and classical Monte Carlo methods~\cite{mc}, and can be measured in 
experiments with cold atoms~\cite{daley-2012,islam-2015,kaufman-2016,
brydges-2018}, unlike the von Neumann entropy. 

R\'enyi and von Neumann entropies are key to understand the fundamental physics 
underlying thermalization in isolated out-of-equilibrium systems, for instance 
after a quantum quench. This is the standard protocol to drive a 
system out-of-equilibrium: A system is prepared initially 
in a low-entanglement state $|\Psi_0\rangle$, and it is let to evolve 
with a local hamiltonian $H$, such that $[H,|\Psi_0\rangle\langle\Psi_0|]
\ne0$. It is by now accepted that for generic systems local properties of the 
steady-state emerging at late times can be described by the Gibbs (thermal) 
ensemble, whereas for integrable models one has to use a Generalized 
Gibbs Ensemble (GGE) \cite{rigol-2007,bastianello-2016,vernier-2016,alba-2015,cc-07,
	cramer-2008,barthel-2008,cramer-2010,calabrese-2011,calabrese-2012,cazalilla-2006,cazalilla-2012a,
	sotiriadis-2012,collura-2013,fagotti-2013,kcc14,sotiriadis-2014,essler-2015,vidmar-2016,pvw-17,
ilievski-2015a,cardy-2015,sotiriadis-2016,mc-12,p-13,fe-13b,fcec-14,langen-15,pk-18}. 
Quite generically, the R\'enyi entropies (and the von Neumann entropy) 
exhibit a linear growth at short times, which is followed by a saturation at 
late times. Their short-time behavior reflects the irreversible growth of 
the complexity of the system after the quench, whereas their steady-state 
value coincides with generalized (GGE) free energies~\cite{alba-2017,
alba-2017-a,mestyan}. These qualitative features, i.e., the linear growth 
and the saturation behavior, appear to be ubiquitous~\cite{calabrese-2005,fagotti-2008,pnas,dmcf-06,ep-08,lauchli-2008,kim-2013,nr-14,coser-2014,cce-15,fc-15,cc-16,chmm-16,buyskikh-2016,kctc-17,mkz-17,r-2017,p-18,fnr-17,bhy-17,hbmr-17,alba-2018,BeTC18,d-17,mbpc-17,BeTC18,nrv-18,bfpc-18,collura-g,nwfs-18}. 

A well-known quasiparticle picture allows to understand the 
qualitative behavior of the entanglement 
spreading~\cite{calabrese-2005}. In this picture the entanglement growth 
is the result of the ballistic propagation of entangled 
pairs of quasiparticles created after the quench. 
For non-interacting fermion models, the quasiparticle picture is exact in the 
scaling limit, as it is has been demonstrated in Ref.~\onlinecite{fagotti-2008}. 
Remarkably, for integrable interacting models, by complementing the 
semiclassical picture with thermodynamic knowledge of the steady 
state, it is possible to describe quantitatively the full-time 
dynamics of the von Neuamann entropy~\cite{pnas,long}. 
Unfortunately, a generalization of this result to the R\'enyi entropies 
is not available yet, although the Thermodynamic Bethe Ansatz (TBA) 
can be used (see Refs.~\onlinecite{alba-2017},\onlinecite{alba-2017-a} and 
\onlinecite{mestyan}) to obtain the steady-state 
R\'enyi entropies.

%
\begin{figure}[t]
\includegraphics*[width=1\linewidth]{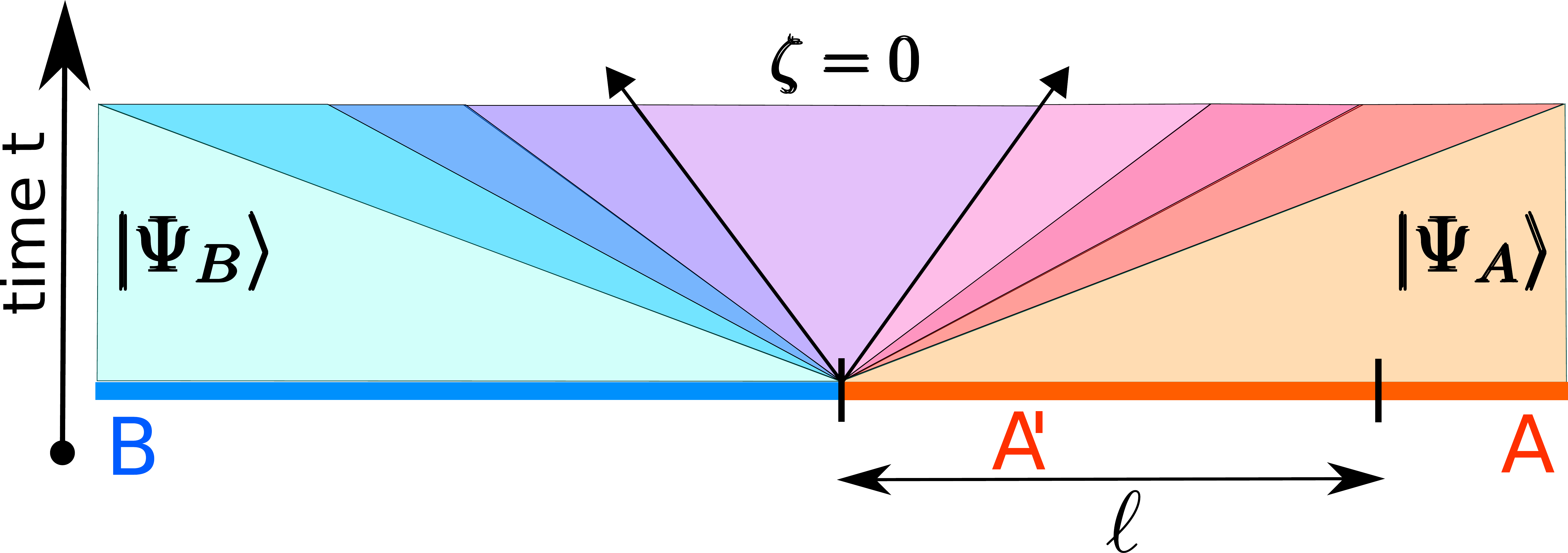}
\caption{ R\'enyi entropies after a  quench from a piecewise 
 homogeneous initial condition in integrable systems: Setup considered in 
 this work. Two chains $A$ and $B$ are prepared in the 
 states $|\Psi_A\rangle$ and $|\Psi_B\rangle$ and are 
 joined together at $t=0$. 
 In the space-time scaling limit on each ray $\zeta$ 
 dynamical properties are described by a Generalized Gibbs 
 ensemble (GGE). 
 Here we focus on the R\'enyi entropies of a subregion 
 $A'$ of length $\ell$ placed next to the interface 
 between $A$ and $B$. 
}
\label{cartoon}
\end{figure}
%

Recent years witnessed an enormous interest in quenches from 
piecewise homogeneous initial conditions. Several techniques 
have been used, such as Conformal Field Theory~\cite{spyros-2008,
bernard-2012,bhaseen-2015,allegra-2016,dubail-2017,dubail-2017a} (CFT), 
free-fermion methods~\cite{eisler-2009,de-luca-2013,sabetta-2013,eisler-2013,alba-2014,collura-2014,de-luca-2016,
eisler-2016,viti-2016,kormos-2017,perfetto-2017,vidmar-2017}, field theory methods~\cite{de-luca-2014,olalla-2014,biella-2016}, 
integrability~\cite{zotos-1997,zotos-1999,prosen-2011,prosen-2013,ilievski-2015}, and 
numerical techniques~\cite{fabian-2003,gobert-2005,langer-2009,karrasch-2012,karrasch-2013,vidmar-2017}. 
For integrable models the recently-developed Generalized Hydrodynamics (GHD)~\cite{olalla-2016,bertini-2016} 
allows for an analytic treatment of these quenches~\cite{doyon-2016,de-luca-2016a,
doyon-2017,doyon-2017a,doyon-2017b,doyon-2017c,bulchandani-2017,bulchandani-2017a,ilievski-2017,doyon-2017d,eb-17,fagotti-null,ilievski-2018,collura-2018,
doyon-2018,alvise-2018,caux-2018,ilievski-2018a}. 
The standard setup that we consider is depicted in Figure~\ref{cartoon}. 
This is the prototypical situation 
to study quantum transport in one-dimensional system. 
Two chains $A$ and $B$ are prepared in two macroscopically different states. 
At $t=0$ they are connected, and the full system is let to evolve unitarily 
under a many-body hamiltonian $H$. 
In the GHD approach, in the scaling limit of long time $t$ 
and large distance $x$, at fixed $\zeta\equiv x/t$ 
(rays in Figure~\ref{cartoon}) dynamical properties are described 
by a GGE. However, as of now, only few results have been 
derived for the entanglement dynamics after quenches from piecewise 
homogeneous states. 
Notable exceptions are systems that can be mapped to CFTs in curved 
spacetime~\cite{allegra-2016,dubail-2017,dubail-2017a}. 
Also, for free-fermion systems, the full-time dynamics of the entanglement 
entropies has been derived analytically in Ref.~\onlinecite{bertini-2018}. 
For interacting integrable models a conjecture for the steady-state 
von Neumann entropy and for the entanglement production rate at short 
times has been presented in Ref.~\onlinecite{alba-2018}. 

However, extending the quasiparticle picture to describe the R\'enyi entropies 
after quenches from piecewise homogeneous states remains an open problem. 
The aim of this work is to provide a step forward in this direction. 
Here we focus on the steady-state R\'enyi entropies 
of a finite subregion $A'$ of length $\ell$ placed next to the interface 
between $A$ and $B$ (see Figure~\ref{cartoon}). The key idea is that 
in the limit $\ell/t\to0$, the entire subsystem $A'$ should be described 
by the GGE with $\zeta=0$, i.e., by the NESS 
(Non Equilibrium Steady State). 
Thus, following Ref.~\onlinecite{alba-2017}, 
the density of R\'enyi entropy has to coincide with that of the NESS 
R\'enyi entropy, which can be calculated using the Thermodynamic Bethe 
Ansatz (TBA). Specifically, the appropriate 
thermodynamic ensemble that determines the R\'enyi entropies is
obtained by combining the GHD approach for the quench~\cite{olalla-2016,bertini-2016} with the results of Ref.~\onlinecite{alba-2017}. By using this method, 
we investigate the R\'enyi entropy after quenches in the spin-$1/2$ 
anisotropic Heisenberg chain (XXZ chain). 
We consider several initial states for $A$ and $B$, such as 
the N\'eel state and the Majumdar-Ghosh state. However, in principle, 
our approach is straightforwardly generalizable to any low-entangled 
initial state (product state) and to any integrable model. 
For all these quenches, we provide robust numerical evidence 
for our results by using the time-dependent 
Density Matrix Renormalization Group (tDMRG) 
method~\cite{white-2004,daley-2004,uli-2005,uli-2011}. 

Finally, we consider the quench in which $A$ is initially 
prepared in the ferromagnetic state. After mapping the Heisenberg chain 
onto a system of interacting fermions confined in a box, 
this corresponds to the trap-expansion protocol that is routinely 
performed in cold-atom experiments. For this reason, we term this quench 
\emph{expansion quench}. 

A remarkable feature of the XXZ chain is that its spectrum 
contains composite excitations, which consist of bound states of 
several elementary particles (magnons). The physics of these bound 
states is receiving constant attention, both theoretically~\cite{hanus-1963,
wortis-1963,fogedby-1980,schneider-1981,ganahl-2012,vl-2015}, 
as well as experimentally~\cite{haller-2009,fukuhara-2013}. 
Interestingly, while for generic quenches, 
the steady-state entropies contain information about all the types 
of quasiparticles (bound and unbound states), this is 
not the case for the expansion quench, at least in the limit 
of large anisotropy. Here we show that this reflects that the bound-state 
transport between the two chain is not possible at large 
anisotropy. This is an intriguing effect of the interactions, which renormalize 
the group velocity of the ``bare'' excitations of the model. Physically, 
after the expansion quench, the velocities of the bound states 
created in the bulk of $A$ and $B$ change their sign when 
approaching the interface, and they are deflected back. 
This suggests that the expansion quench could be used as 
a filter to isolate multi-spin bound states. 
An interesting result is that there is a ``critical'' 
anisotropy, below which dynamical properties change abruptly, and 
the bound-state transport is permitted. We explicitly check 
this scenario for the expansion of the N\'eel state and the 
Majumdar-Ghosh state, although we expect it to happen  
for a larger class of initial states. 

The paper is organized as follows. In section~\ref{sec-TBA} and~\ref{sec-tba-quench} 
we review the Thermodynamic Bethe ansatz (TBA) approach for homogeneous 
quantum quenches in integrable models, focusing on the steady-state R\'enyi entropies 
in section~\ref{sec-tba-renyi}. In section~\ref{sec-ghd} we review the 
Generalized Hydrodynamic approach (GHD). In section~\ref{sec-ghd-main} 
we discuss the steady-state R\'enyi entropies. 
In section~\ref{sec-model} we present the XXZ chain and the quench protocols. 
In section~\ref{sec-ghd-eq} and~\ref{sec-ghd-eq1} we discuss the 
expansion quench and its GHD solution, highlighting the absence of bound-state 
transport for large anisotropy. In 
section~\ref{sec-ghd-eq2} we report some exact results for the expansion quenches 
of the N\'eel and the Majumdar-Ghosh state in the limit of large anisotropy. 
In section~\ref{sec-eq-renyi} we present theoretical predictions for 
the steady-state R\'enyi entropies after the expansion quench. Numerical 
results are presented in section~\ref{sec-xxz-num}. 
Finally, these are benchmarked against tDMRG simulations in 
section~\ref{sec-numerics}. 

\section{R\'enyi entropies after a homogeneous quench} 
\label{sec-renyi-gge}

Here we summarize the Thermodynamic Bethe Ansatz (TBA) approach introduced 
in Ref.~\onlinecite{alba-2017} (see also Ref.~\onlinecite{alba-2017-a} 
and~\onlinecite{mestyan}) 
to calculate the steady-state R\'enyi entropies after a homogeneous quench. 
First, in section~\ref{sec-TBA} we 
review some general aspects of the TBA approach for integrable models. 
In section~\ref{sec-tba-quench} we discuss the TBA treatment for 
quenches. Section~\ref{sec-tba-renyi} is devoted to discussing  
how to calculate the entropies using TBA. 

\subsection{Thermodynamic Bethe ansatz (TBA) for integrable models}
\label{sec-TBA}

The distinctive feature of 
integrable models is that they possess families of well-defined and 
stable, i.e., having infinite lifetime, quasiparticles. Different 
quasiparticles are labelled by their quasimomentum (rapidity) that we 
denote as $\lambda$. A generic eigenstate of a Bethe ansatz solvable 
model is in a one-to-one correspondence with a set of allowed 
quasiparticles rapidities, which are obtained by solving a system of nonlinear 
algebraic equations known as Bethe equations~\cite{taka-book}. In the 
thermodynamic limit the rapidities form a continuum, and it is more convenient 
to work with the rapidity density $\pmb{\rho}$ (particle densities) and 
the hole densities $\pmb{\rho}^{\scriptscriptstyle(h)}$, i.e., the 
density of unoccupied rapidities. Here we defined $\pmb{\rho}\equiv
\{\rho_n\}_{n=1}^{\cal N}$, and $\pmb{\rho}^{\scriptscriptstyle(h)}\equiv
\{\rho_n^{\scriptscriptstyle(h)}\}_{n=1}^\infty$, where $n$ labels the 
different quasiparticles families. Their total number 
${\cal N}$ depends on the model under 
consideration. For instance, in the XXZ chain with $\Delta\ge 1$, 
which is the model considered here, there are infinite families of 
quasiparticles, i.e., ${\cal N}=\infty$. 
For the XXZ chain the quasiparticles with 
$n=1$ correspond to free magnon-like excitations, whereas 
the ones with $n>1$ are bound states of $n$ down spins. 
For later convenience we also 
define the total densities $\rho_n^{\scriptscriptstyle(t)}$, 
the densities $\eta_n$, and the filling functions $\vartheta_n$ as 
\begin{align}
\label{d1}
& \rho_n^{\scriptscriptstyle (t)}\equiv \rho_n+\rho^{\scriptscriptstyle (h)}_n,\\
\label{d2}
& \eta_n\equiv\rho_n^{\scriptscriptstyle(h)}\rho_n^{-1},\\
\label{d3}
& \vartheta_n\equiv[1+\eta_n]^{-1}. 
\end{align}
In any Bethe ansatz solvable model, the particle densities $\rho_n$ 
are coupled to the $\eta_n$ via the thermodynamic version of the Bethe 
equations, which read~\cite{taka-book}  
\begin{equation}
\label{tba-eq}
2\pi\rho_n(1+\eta_n)=a_n-\sum_{m=1}^\infty a_{nm}\star \rho_m. 
\end{equation}
Here, the functions $a_n(\lambda)$ and $a_{nm}(\lambda)$ are known 
from the Bethe ansatz solution of the model. 
For the XXZ chain $a_n$ and $a_{nm}$ are given as~\cite{taka-book} 
\begin{align}
& a_n=\frac{1}{\pi}\frac{\sinh(n\eta)}{\cosh(n\eta)-\cos(2\lambda)}\\
& a_{nm}=(1-\delta_{nm})a_{|n-m|}+a_{|n-m|+2}+\\\nonumber
 &\quad\cdots+a_{n+m-2}+a_{n+m}. 
\end{align}
Here $\eta\equiv\textrm{arccosh}(\Delta)$. 
The matrix $a_{nm}$ encodes the scattering between 
quasiparticles of different families, and with different 
rapidities. In~\eqref{tba-eq} the symbol $\star$ denotes the 
convolution
\begin{equation}
f\star g\equiv\int d\mu f(\lambda-\mu)g(\mu). 
\end{equation}
Any set of densities $\pmb{\rho}$ identifies a thermodynamic 
macrostate, which corresponds to an exponentially diverging 
(with size) number of microscopic eigenstates. 
Any of these eigenstates can be chosen as a finite-size 
representative of the thermodynamic macrostate. The 
total number of equivalent microstates is given as $e^{S_{YY}}$, 
with $S_{YY}$ the so-called Yang-Yang entropy~\cite{taka-book} 
\begin{multline}
\label{s-yy}
S_{YY}(\pmb{\rho})\equiv L\sum_n\int d\lambda \{
\rho^{(t)}_n\ln\rho_n^{(t)}\\-\rho_n^{(h)}\ln\rho_n^{(h)}-\rho_n\ln\rho_n\}, 
\end{multline}
where $L$ is the system size. The Yang-Yang entropy is extensive and it 
is obtained by summing independently over the quasiparticle families and 
their rapidity, reflecting, again, the integrability. 
For systems in thermal equilibrium 
$S_{YY}$ is the thermal entropy. For out-of-equilibrium systems, the 
density of the 
Yang-Yang entropy of the GGE that describes the steady state is a key ingredient 
to understand quantitatively the dynamics of the entanglement entropy 
after the quench~\cite{pnas,long}.

\subsection{TBA description of homogeneous quenches}
\label{sec-tba-quench}

Integrable models {\it \`a la} Bethe possess an 
extensive number of local and quasilocal conserved quantities 
$\hat Q_k$, i.e., having the property that $[\hat Q_j,\hat Q_k]=0,\,
\forall j,k$, with $\hat Q_2$ being the hamiltonian. 
This implies that, due to these conservation laws, in integrable 
models the out-of-equilibrium dynamics after 
a quantum quench, and the steady state, are strongly constrained. 
As a consequence, at late times the system fails to thermalize, 
i.e., {\it local} properties are not described by the Gibbs ensemble. 
Still, it is now accepted that local and quasilocal properties of the steady state 
are described by a Generalized Gibbs Ensemble (GGE). The GGE density matrix 
is obtained by complementing the Gibbs density matrix with the extra conserved 
quantities $\hat Q_j$, to obtain  
\begin{equation}
\label{rho-gge}
\rho_{\textrm{GGE}}=\frac{1}{Z_{\textrm{GGE}}}
\exp\Big(-\sum_k\beta_k \hat Q_k\Big). 
\end{equation}
Here $\beta_k$ is the Lagrange multiplier associated with 
$\hat Q_k$ and $Z_{\textrm{GGE}}$ is a normalization constant (GGE 
partition function). The $\beta_k$ are fixed by imposing that the 
GGE average of $\hat Q_k$ equals their initial state value as 
\begin{equation}
\label{constrgge}
\textrm{Tr}(\rho_{\textrm{GGE}}\hat Q_k)=\langle\Psi_0|\hat 
Q_k|\Psi_0\rangle. 
\end{equation}
The key idea of the TBA approach for quantum quenches is that GGE 
expectation values of local and quasilocal observables are 
described by a carefully chosen thermodynamic macrostate 
(see section~\ref{sec-TBA}), which can be completely characterized 
in terms of the initial state expectation values of the 
conserved quantities~\cite{iqdb-15} (cf.~\eqref{constrgge}). 

We now illustrate how to determine this macrostate by using 
the TBA. First, in the thermodynamic limit, the expectation value 
$\hat Q_k$ over a generic thermodynamic 
macrostate identified by particle and hole distributions 
$\rho_n,\rho_n^{\scriptscriptstyle(h)}$ is obtained by 
summing over the quasiparticle families and integrating 
over their rapidities as~\cite{taka-book} 
\begin{equation}
\label{charges}
Q_k=L\sum_n\int d\lambda f_{kn}(\lambda)\rho_n(\lambda), 
\end{equation}
where we introduced the functions $f_{kn}$, and $L$ is the system size. 
It is convenient to define the generalized driving $g_n$ as 
\begin{equation}
\label{gn-def}
g_n(\lambda)\equiv\sum_k\beta_k f_{kn}. 
\end{equation}
The driving $g_n$ contains the crucial information about the initial 
values of the conserved charges, or, equivalently, on $\beta_k$. 

To proceed, it is useful to consider the GGE expectation value 
$\textrm{Tr}\rho_{\textrm{GGE}}\hat {\mathcal O}$ of a generic local (or 
quasilocal) observable $\hat {\mathcal O}$. In the thermodynamic limit 
the trace over the model eigenstates becomes a functional integral as 
\begin{equation}
\label{sum}
\textrm{Tr}\rightarrow\int D{\rho} e^{S_{YY}({\rho})}, 
\end{equation}
Here $D{\rho}\equiv\prod_n D\rho_n$ denotes the functional 
integral over the densities $\rho_n$. The Yang-Yang entropy in~\eqref{sum} 
takes into account that a thermodynamic macrostate corresponds to 
an exponentially large number of eigenstates. Using~\eqref{sum}, 
one can write 
\begin{equation}
\label{gge-eq}
\textrm{Tr}(\hat{\mathcal O}\rho_{\textrm{GGE}})=
\frac{1}{Z_{\textrm{GGE}}}\int D{\rho}e^{ -{\cal E}({\rho})
+S_{YY}({\rho})}
\langle{\rho}|\hat {\mathcal O}|{\rho}\rangle. 
\end{equation}
Here we defined $\cal{E}({\rho})$ as 
\begin{equation}
\label{eps}
{\cal{E}}=L\sum_{n}\int d\lambda g_{n}(\lambda)\rho_n(\lambda). 
\end{equation}
In~\eqref{gge-eq}, $\langle{\rho}|\hat{\mathcal O}|{\rho}\rangle$ 
denotes the value of $\hat{\mathcal O}$ over the macrostate. 
Importantly, the locality and quasilocality of $\hat Q_k$ implies that 
${\cal E}$ is extensive, as it is clear from~\eqref{eps}. 
Due to the extensivity of ${\cal E}$ and $S_{YY}$, following the 
standard TBA approach~\cite{taka-book}, in the thermodynamic limit the integral 
in~\eqref{gge-eq} can be treated using the saddle point method. 
One has to minimize the functional ${\cal S}_{\textrm{GGE}}$ 
defined as 
\begin{equation}
\label{gge-f}
{\cal S}_{\textrm{GGE}}\equiv -{\cal E}+S_{YY}. 
\end{equation}
After minimizing~\eqref{gge-f} with respect to $\rho_n$, one obtains 
a set of generalized TBA equations for $\eta_n$ (see~\eqref{d3})as  
\begin{equation}
\label{s-eta}
\ln\eta_n=g_n+\sum_{m=1}^\infty a_{nm}\star\ln(1+\eta_m^{-1}(\lambda)), 
\end{equation}
where $a_{nm}$ is the same as in~\eqref{tba-eq}, and $g_n$ is defined 
in~\eqref{gn-def}. Finally, the GGE macrostate densities $\rho_n$ 
are obtained by substituting the solutions of~\eqref{s-eta} in 
the TBA equations~\eqref{tba-eq}. 

\subsection{TBA approach for R\'enyi entropies}
\label{sec-tba-renyi}

Here we discuss how to calculate the steady-state value of the R\'enyi 
entanglement entropies after a homogeneous quench in integrable 
systems~\cite{alba-2017,alba-2017-a}. Since local properties of the 
steady state are described by a GGE, the density of R\'enyi entanglement 
entropies has to coincide with that of the GGE R\'enyi entropies 
$S_{\textrm{GGE}}^{\scriptscriptstyle(\alpha)}$. These are 
defined as 
\begin{equation}
\label{renyi-gge-def}
S_{\textrm{GGE}}^{\scriptscriptstyle(\alpha)}=\frac{1}{1-\alpha}\ln\textrm{Tr}
\rho_{\textrm{GGE}}^\alpha, 
\end{equation}
where $\rho_{\textrm{GGE}}$ is defined in~\eqref{rho-gge}. 
We now discuss how to calculate~\eqref{renyi-gge-def} using the Thermodynamic 
Bethe ansatz. Similar to~\eqref{gge-eq}, in the thermodynamic limit 
the GGE R\'enyi entropies are written as 
\begin{equation}
\label{gen-qa}
S_{\textrm{GGE}}^{(\alpha)}=\frac{1}{1-\alpha}\Big[
\ln\int D\rho\exp(-\alpha{\cal{E}}+S_{YY})+\alpha f_{\textrm{GGE}}\Big], 
\end{equation}
where $f_{\textrm{GGE}}\equiv -\ln Z_{\textrm{GGE}}$, 
${\cal{E}}$ is the same as in~\eqref{eps}, and $S_{YY}$ is 
the GGE thermodynamic entropy. 
The functional integral in~\eqref{gen-qa} can be treated using 
the saddle point method. The functional ${\cal S}_{
\textrm{\textrm{GGE}}}^{(\alpha)}$ that has to be minimized 
depends explicitly on the R\'enyi index $\alpha$, and it is 
defined as 
\begin{equation}
\label{r-fun}
{\cal S}_{\textrm{\textrm{GGE}}}^{(\alpha)}\equiv-
\alpha{\cal E}+S_{YY}. 
\end{equation}
The saddle point condition leads to the modified TBA equations for 
$\eta_{n}^{\scriptscriptstyle(\alpha)}$ as 
\begin{equation}
\label{m-eta}
\ln\eta^{(\alpha)}_n=\alpha g_n+\sum_{m=1}^\infty a_{nm}\star\ln(1+
1/\eta^{(\alpha)}_m(\lambda)). 
\end{equation}
The particle densities $\rho_{n}^{\scriptscriptstyle(\alpha)}$ are 
obtained by substituting $\eta_n^{\scriptscriptstyle(\alpha)}$ in 
the Bethe equations~\eqref{tba-eq}. 
By combining~\eqref{r-fun} and~\eqref{renyi-gge-def}, one obtains that 
the density of the R\'enyi entropies for a GGE is 
\begin{equation}
\label{main}
S_{\textrm{GGE}}^{(\alpha)}=\frac{1}{1-\alpha}\Big[
\left.\Big(-\alpha{\cal E}+S_{YY}\Big)\right|_{\rho_n^{(\alpha)}}
+\alpha \left.f_{\textrm{GGE}}\right|_{\rho_n^{(1)}}\Big], 
\end{equation}
where $\rho_n^{(1)}$ is the saddle point 
densities for $\alpha=1$ that describes local steady-state properties. 

As it is clear from~\eqref{m-eta}, the macrostate describing 
the R\'enyi entropies depends on $\alpha$. This is surprising because 
both the R\'enyi entropies and the von Neumann entropy are 
calculated from the same quantum state. An interesting consequence is 
that the region of energy spectrum of the post-quench hamiltonian that is 
relevant for describing the R\'enyi entropies is different from that 
describing the local observables. A similar behavior happens for 
non-integrable models, where it is a consequence of the Eigenstate 
Thermalization Hypothesis~\cite{garrison-2015}. 

At this point it is important to stress that the method to calculate 
the R\'enyi entropies that we outlined so far requires as crucial ingredient 
the set of infinitely many Lagrange multipliers $\beta_{k}$ entering 
in the driving functions $g_n$ (cf.~\eqref{gn-def}). 
Fixing the $\beta_k$ by using the constraint~\eqref{constrgge}  is 
a formidable task that cannot be carried out in practice.  
However, this difficulty can be overcome in two ways. One is to use 
the  Quench Action method~\cite{caux-2013,caux-2016}. 
The Quench Action provides direct access to the thermodynamic macrostate 
describing the stationary state. The driving 
functions $g_n$ are extracted from the overlaps between the initial 
state and the eigenstates of the post-quench hamiltonian, without 
relying on the $\beta_k$. Interestignly, a subset of the 
thermodynamically relevant overlaps is sufficient to determine 
the macrostate (see~\cite{wdbf-14,PMWK14,ac-16qa}). 
For a large class of initial states the overlaps can be 
determined analytically~\cite{fcc-09,Pozsgay18,grd-10,pozsgay-14,dwbc-14,bdwc-14,cd-12,pc-14,fz-16,dkm-16,hst-17,msca-16,pe-16,BeSE14,BePC16,BeTC17}. 
This holds also for systems in the continuum, such as the Lieb-Liniger gas. 
For instance, in Refs.~\cite{dwbc-14,bdwc-14,cd-12,dwbc-14,pce-16}, 
the overlaps between the Bose condensate (BEC) state and the 
eigenstates of the Lieb-Liniger model have been calculated, for both 
attractive and repulsive interactions. 
Interestingly, a crucial feature of all the initial states for which 
it has been possible to calculate the overlaps is reflection symmetry. 
Reflection-symmetric initial states have nonzero overlap only 
with parity-invariant eigenstates, which are identified by 
solutions of the Bethe equations containing only pairs of rapidities 
with opposite sign, i.e., such that $\{\lambda_j\}=\{-\lambda_j\}$. 
Interestingly, the role played by parity-invariance for the 
solvability of quantum quenches has been investigated in 
Ref. \cite{ppv-17} for lattice models and in Ref. \cite{delfino-14}  
for integrable field theories. 
Also, one should remark that non-reflection-symmetric 
initial states have been found~\cite{BeTC17} 
for which the Quench Action method can be applied.
We should mention that the knowledge of the overlaps 
was crucial in Ref.~\cite{alba-2017-a} 
to obtain the steady-state R\'enyi entropies after the 
quench from the N\'eel state. 
For some quenches the TBA macrostates describing the 
steady state can be derived from the expectation values of the 
conserved quantities over the initial state, without knowing 
the overlaps~\cite{iqdb-15}. For instance, this is the case for the quench 
from the tilted N\'eel state~\cite{pvc-16} (see, however, 
Ref.~\onlinecite{Pozsgay18} for a recent conjecture for the 
overlaps). In these cases, one can use the TBA 
equations~\eqref{s-eta} to extract the driving 
functions $g_n$ from the macrostate densities. The functions $g_n$ 
can then be used in the generalized 
TBA equations~\eqref{m-eta} to derive the R\'enyi entropies. 
The validity of this approach for homogeneous quenches has 
been investigated recently in Ref.~\onlinecite{mestyan}.  
Crucially, for quenches from piecewise homogeneous initial 
states, the GGE macrostate describing the local steady-state 
are determined by solving a continuity equation 
(see below), and the TBA driving functions are not 
even defined. This implies that the approach of 
Ref.~\onlinecite{mestyan} has to be employed.

\subsection{Min entropy of the GGE}
\label{sec-ghd-sc}

Before proceeding, it is interesting to consider the limit $\alpha\to\infty$ of the 
GGE R\'enyi entropies, which corresponds to min entropy of the GGE, or the 
so-called single copy entanglement entropy $S^{\scriptscriptstyle(
\infty)}$, which is written in terms of the largest eigenvalues $\lambda_M$ 
of the reduced density matrix as $S^{\scriptscriptstyle(\infty)}=-
\ln\lambda_M$. For homogeneous quenches in the XXZ chain the 
steady-state value of the single-copy entanglement has been 
derived in Ref.~\onlinecite{alba-2017}. An interesting feature 
is that the thermodynamic macrostate that determines its 
steady-state value has zero Yang-Yang entropy (see 
Ref.~\onlinecite{mestyan} for a discussion of some general conditions 
for this to hold true). 
Moreover, for quenches in the strong anisotropy limit of the 
XXZ chain (see section~\ref{sec-model} for its definition) 
this macrostate is the ground state of the chain. Notice that 
the ground state of the XXZ chain does not contain bound states. 
This picture breaks down at sufficiently small values of 
the chain anisotropy, where $S^{\scriptscriptstyle(\infty)}$ 
is described by a different macrostate. This new 
macrostate has zero Yang-Yang entropy and the associated TBA 
densities $\rho^{\scriptscriptstyle(\infty)}_n$ exhibit a 
Fermi-sea structure~\cite{alba-2017}, but it has a 
non-trivial bound-state content.  

We now discuss the TBA derivation of the steady-state value of the 
min entropy after a homogeneous 
quenches. The generalization to quenches from piecewise homogeneous 
initial states is straightforward. First, it is useful to rewrite 
the TBA densities $\eta_n^{\scriptscriptstyle(\alpha)}$ 
(cf.~\eqref{m-eta}) as~\cite{alba-2017-a} 
\begin{equation}
\label{ansatz}
\eta_n^{(\alpha)}=\exp(\alpha\gamma_n). 
\end{equation}
Here the functions $\gamma_n$ have to be determined by using 
the TBA equations~\eqref{m-eta}. Clearly, from~\eqref{ansatz} one 
has that in the limit $\alpha\to\infty$, $\eta_n^{\scriptscriptstyle
(\alpha)}$ (vanish) diverge  for ($\gamma_n<0$) $\gamma_n>0$. 
Using~\eqref{ansatz}, the TBA equations~\eqref{m-eta} for 
$\eta_n^{\scriptscriptstyle(\alpha)}$ now read 
\begin{multline}
\label{integ}
\alpha\gamma_n=\alpha g_n\\
+\sum_m\int d\mu a_{nm}(\mu-\lambda)\ln[1+\exp(-\alpha\gamma_m(\mu))], 
\end{multline}
In the limit $\alpha\to\infty$, the last term in~\eqref{integ} 
in nonzero only for $\gamma_m<0$, and it becomes linear in $\alpha$. 
This allows one to rewrite~\eqref{integ} as  
\begin{equation}
\label{integ1}
\gamma_n=g_n+\sum_m\int d\mu a_{nm}(\mu-\lambda)
\gamma_m^+(\mu), 
\end{equation}
where we defined $\gamma_n^+$ as 
\begin{equation}
\label{constr}
\gamma_n^+\equiv\left\{\begin{array}{cc}
\gamma_n & \textrm{if}\,\gamma_n<0,\\
0 & \textrm{if}\,\gamma_n>0.
\end{array}\right.
\end{equation}
Notice that, as it is clear from the definition~\eqref{constr}, the 
integral equations in~\eqref{integ1} are nonlinear. After solving~\eqref{integ1}, the 
particle densities $\rho_n^{\scriptscriptstyle(\infty)}$ are obtained by 
using the TBA equations~\eqref{tba-eq}.

One obtains that $\rho^{\scriptscriptstyle (\infty)}_n$ are 
nonzero only for $\lambda$ such that $\gamma_n<0$. This happens  
because the $\rho_n^{\scriptscriptstyle(\infty)}$ 
cannot diverge, whereas from~\eqref{ansatz}, one has 
that $\eta_n^{\scriptscriptstyle(\infty)}$ diverge for $\lambda$ 
such that $\gamma_n>0$. More quantitatively, in the limit $\alpha\to\infty$ 
the system~\eqref{tba-eq} becomes  
\begin{equation}
\label{rho-in}
2\pi\rho_n=a_n
-\sum\limits_{m=1}^\infty
\int d\mu a_{nm}(\mu-\lambda)\rho_m(\mu)\theta_H(-\gamma_m),
\end{equation}
where, again, $\rho_n$ is nonzero only for $\lambda$ such that 
$\gamma_n<0$. The equation for the hole density 
$\rho_n^{\scriptscriptstyle(h)}$ is given as 
\begin{equation}
\label{rhoh-in}
2\pi\rho^{(h)}_n=a_n-\sum\limits_{m=1}^\infty
\int d\mu a_{nm}(\mu-\lambda)\rho_m(\mu)\theta_H(-\gamma_m), 
\end{equation}
Notice that~\eqref{rhoh-in} is the same as~\eqref{rho-in}, although 
the support, i.e., the values of $\lambda$ for which $\rho_n$  and 
$\rho_n^{\scriptscriptstyle(h)}$ are nonzero, is different. 
In fact, the supports of particle and hole densities are  
complementary. By using~\eqref{s-yy}, it is straightforward to 
check that this implies that the Yang-Yang entropy is zero. 
Finally, by taking the limit $\alpha\to\infty$ in~\eqref{main}, one obtains 
that  
\begin{equation}
\label{sc-ent}
S^{(\infty)}_{\textrm{GGE}}=\left.{\cal E}\right|_{\rho_n^{(\infty)}}
+L \left.f_{\textrm{GGE}}\right|_{\rho_n^{(1)}},  
\end{equation}
where $\rho_n^{(1)}$ is the macrostate describing (quasi) 
local observables in the steady state. 
Apart from the contribution of the GGE grand-canonical 
potential $f_{\textrm{GGE}}$, the min entropy is determined by 
the driving ${\cal E}$ only. Also, for quenches that can be 
treated with the Quench Action, one has $f_{\textrm{GGE}}=0$, 
due to the normalization of the overlaps~\cite{wdbf-14}, 
which further simplifies~\eqref{sc-ent}.

\section{Obtaining the R\'enyi entropies in GHD} 
\label{sec-ghd-renyi}

Here we discuss the GHD approach to calculate the steady-state 
R\'enyi entropies in section~\ref{sec-ghd} 
(see Refs.~\onlinecite{olalla-2016,bertini-2016}), focusing on the entropies in 
section~\ref{sec-ghd-main}. 

\subsection{Generalized Hydrodynamics (GHD)}
\label{sec-ghd}

Let us consider the setup that is depicted in Figure~\ref{cartoon}. 
After the quench, information spreads ballistically from the interface 
between $A$ and $B$. At long times, local and 
quasilocal quantities depend only on the ratio $\zeta=x/t$, with $t$ 
the time after the quench and $x$ the distance from the interface 
between $A$ and $B$. A remarkable result is that in the scaling 
limit of long times and large distances, for each fixed 
ray $\zeta$, dynamical properties of the system are described by a 
GGE~\cite{olalla-2016,bertini-2016}. Similar to homogeneous quenches, this GGE is identified by a 
thermodynamic macrostate (see section~\ref{sec-tba-quench}), i.e., 
by a set of particle and hole densities $\rho_{\zeta,n}$ and 
$\rho_{\zeta,n}^{\scriptscriptstyle(h)}$. For $\zeta=0$ the 
macrostate identifies the so-called Non Equilibrium Steady State 
(NESS). We anticipate that the NESS is the relevant macrostate for 
describing the steady-state R\'enyi entropies of a 
subsystem placed at the interface between $A$ and $B$.  
Clearly, deep in the bulk of the two chains, i.e., for 
$x/t\to\pm\infty$, the steady state is the same as that arising 
after the homogeneous quenches with initial states $|\Psi_A\rangle$ 
and $|\Psi_B\rangle$. Let us denote as $\rho_{\pm\infty,n}$ 
the macrostates that describe these steady states. 

The central result of the Generalized Hydrodynamics~\cite{olalla-2016,bertini-2016} 
is that for generic $\zeta=x/t$, the macrostate $\rho_{\zeta,n}$ 
satisfies a continuity equation having $\rho_{\pm\infty,n}$ as 
boundary conditions for $\zeta\to\pm\infty$. The continuity 
equation is conveniently written in terms of the filling 
functions $\vartheta_{\zeta,n}$ (cf.~\eqref{d3}) 
as  
\begin{equation}
\label{cont}
[\zeta-v_{\zeta,n}(\lambda)]\partial_\zeta\vartheta_{\zeta,n}(\lambda)=0, 
\end{equation}
where $v_{\zeta,n}$ are the group velocities of the low-lying (particle-hole) 
excitations around ${\rho}_{\zeta,n}$. 
Crucially, in~\eqref{cont} there is no explicit 
scattering term between different quasiparticles. This is a consequence 
of integrability. Indeed, for interacting integrable models, the full 
effect of interactions is taken into account by the renormalization 
of the group velocity $v_{\zeta,n}$. The TBA calculation of $v_{\zeta,n}$ 
will be illustrated below. 
Notice that in Eq.~\eqref{cont} each ray $\zeta$ can be treated separately. 
The general solution of~\eqref{cont} can be written as~\cite{olalla-2016,bertini-2016} 
\begin{equation}
\label{ghd-sol}
\vartheta_{\zeta,n}(\lambda)=\theta_H(v_{\zeta,n}-\zeta)(
\vartheta_{+\infty,n}-\vartheta_{-\infty,n})+\vartheta_{-\infty,n}. 
\end{equation}
Here $\theta_H$ is the Heaviside theta function and $\vartheta_{\pm\infty,n}$ 
are the boundary conditions in the limits $\zeta\to\pm\infty$. 
Again, they identify the macrostates describing the steady 
state after the quenches from $|\Psi_A\rangle$ 
and $|\Psi_B\rangle$ (see Figure~\ref{cartoon}), respectively. 
Notice that the solution~\eqref{ghd-sol} is only implicit 
because the velocities $v_{\zeta,n}$ have to be determined self 
consistently. 
The equations that determine the particle density $\rho_{\zeta,n}$ 
are the same as in~\eqref{tba-eq}. In terms of $\vartheta_{\zeta,n}$ 
they read as  
\begin{equation}
	2\pi\rho_{\zeta,n}\vartheta_{\zeta,n}^{-1}=a_n-\sum_{m=1}^\infty a_{nm}\star 
\rho_{\zeta,m}, 
\end{equation}
where $\vartheta_{\zeta,n}$ are obtained from~\eqref{ghd-sol}. 

As anticipated, the crucial ingredient of the GHD are the group velocities 
$v_{\zeta,n}$. 
In the Bethe ansatz framework, these are constructed as particle-hole 
excitations above ${\rho}_{\zeta,n}$. 
For generic integrable {\it interacting} models the effect of 
particle-hole excitations is to renormalize (``dress'') the 
quasiparticle rapidities. This is reflected in a renormalization of 
the quasiparticle energies and group velocities. 

In the following we describe how to calculate this renormalization by using 
the approach described in Ref.~\onlinecite{bonnes-2014}. 
We denote the ``dressed'' single particle energies by $e_{\zeta,n}$, 
whereas the ``bare'' ones are denoted by $\epsilon_n$. 
In terms of $e_{\zeta,n}$, the group velocities $v_{\zeta,n}$ are written 
as~\cite{bonnes-2014} 
\begin{equation}
\label{group-v}
v_{\zeta,n}=\frac{e'_{\zeta,n}}{2\pi\rho^{\scriptscriptstyle (t)}_{\zeta,n}}.
\end{equation}
Here $e_{\zeta,n}'\equiv d e_{\zeta,n}/d\lambda$, and $\rho_n^{\scriptscriptstyle
(t)}$ is the total density of the thermodynamic macrostate. 
The functions $e_{\zeta,n}'$ are obtained by solving 
the system of nonlinear integral equations~\cite{bonnes-2014} 
\begin{multline}
\label{v-int}
e_{\zeta,n}'-\epsilon_{n}'\\+\frac{1}{2\pi}\sum_m\int d\mu e'_{\zeta,m}
(\mu)a_{nm}(\mu-\lambda)\vartheta_{\zeta,m}(\mu)=0.
\end{multline}
Here $a_{nm}$ is the same scattering matrix as in the TBA equations~\eqref{tba-eq}, and 
we defined $\epsilon'_{n}\equiv d\epsilon_{n}/d\lambda$. 
A very efficient strategy to solve~\eqref{cont} and~\eqref{v-int} is 
by iteration: One starts with an initial guess for $v_{\zeta,n}$, 
using~\eqref{ghd-sol} to derive $\vartheta_{\zeta,n}$. Thus, a new 
set of velocities is determined by using~\eqref{group-v} and~\eqref{v-int}. 
These two steps are iterated until convergence is reached.

\subsection{GHD approach for R\'enyi entropies}
\label{sec-ghd-main}

Here we consider a finite 
subsystem $A'$ of length $\ell$ at the interface between $A$ and $B$ 
(see Figure~\ref{cartoon}). We are interested in the 
steady-state value of the R\'enyi entropy density $S^{\scriptscriptstyle(\alpha)}/
\ell$ of $A'$ for generic $\alpha$. For $\alpha=1$ the result has been 
provided in Ref.~\onlinecite{alba-2017}
The key idea is that for any finite $\ell$, in the limit $t\to\infty$ 
the density of R\'enyi entropy has to coincide with that of the 
R\'enyi entropy of the GGE that describes the local 
equilibrium state. Now, in the limit $t\to\infty$ any finite region 
around the interface between $A$ and $B$ is described by the GGE with 
$\zeta=0$. 

The general idea to obtain the steady-state R\'enyi entropies 
is to combine the GHD framework (see section~\ref{sec-ghd}) 
with the TBA method for the R\'enyi entropies (see 
section~\ref{sec-tba-renyi}). 
The densities $\vartheta_{\zeta=0,n}$ 
are obtained by solving the GHD continuity equations~\eqref{cont}.  
Then, the R\'enyi entropies are obtained using the TBA method described 
in section~\ref{sec-renyi-gge}. Crucially, the second step requires 
knowing the driving $g_n$ (see~\eqref{s-eta} for its definition) that 
determines $\eta^{\scriptscriptstyle(\alpha)}_n$. 
In contrast with homogeneous quenches, $g_n$ is not 
known a priori. Here, following Ref.~\onlinecite{mestyan} $g_n$ is extracted from the 
TBA equations~\eqref{s-eta}, by substituting the densities 
$\vartheta_{\zeta=0,n}$ obtained from~\eqref{cont}. 
Finally, $g_n$ is used in the TBA equations~\eqref{m-eta} 
and~\eqref{tba-eq} to obtain the densities 
$\eta_n^{\scriptscriptstyle(\alpha)}$ and 
$\rho_{n}^{\scriptscriptstyle(\alpha)}$. The final expression 
for the R\'enyi entropies densities is given 
by~\eqref{main}. 

\section{Analytical results for the XXZ chain}
\label{sec-xxz} 

In this section, by using the approach presented in section~\ref{sec-ghd-renyi}, 
we calculate the steady-state R\'enyi entropies 
after a quench from a piecewise homogeneous initial state in the XXZ chain. 
Section~\ref{sec-model} introduces the XXZ and  the quench protocol,  
in particular, the {\it expansion quench}. 
The GHD solution of the expansion quench is detailed in  
section~\ref{sec-ghd-eq},~\ref{sec-ghd-eq1},~\ref{sec-ghd-eq2}. 
Finally, in section~\ref{sec-eq-renyi} we present 
our analytical predictions for the entropies.  

\subsection{Model and quench protocols (expansion quench)} 
\label{sec-model}

The spin-$1/2$ XXZ chain is defined by the hamiltonian 
\begin{equation}
\label{xxz-ham}
H=\sum\limits_{i=1}^L\Big[\frac{1}{2}(S_i^+S_{i+1}^-+S_i^-S_{i+1}^-)+\Delta S_i^zS_{i+1}^z\Big], 
\end{equation}
where $S_i^{+,-,z}$ are spin-$1/2$ operators and $\Delta$ is the 
anisotropy parameter. We consider periodic boundary conditions by 
identifying sites $1$ and $L+1$ of the chain. We restrict 
ourselves to $\Delta>1$. 
In the generic quench protocol, at $t=0$ the two chains $A$ and $B$ 
(see Figure~\ref{cartoon}) are prepared in two macroscopically 
different states $\Psi_A$ and $\Psi_B$. At $t>0$ the unitary dynamics 
under the XXZ hamiltonian~\eqref{xxz-ham} is investigated. 
Here we also consider a special type of quench, which we 
term {\it expansion quench}. In the expansion quench 
part $A$ (see Figure~\ref{cartoon}) is prepared in the vacuum 
state of the quasiparticles, which for the XXZ chain is the 
ferromagnetic state with all the spins pointing up, i.e., 
\begin{equation}
|F\rangle\equiv\left|\uparrow\uparrow\uparrow\cdots\right\rangle. 
\end{equation}
After mapping the XXZ chain onto a system of interacting fermions, the 
expansion quench is equivalent to a box-trap expansion, which is routinely 
used in cold-atom experiments. 
Here we only consider the expansion of the N\'eel state $|N\rangle$ and 
of the Majumdar-Ghosh (or dimer) state $|MG\rangle$. The N\'eel state is 
defined as 
\begin{equation}
|N\rangle\equiv\frac{1}{\sqrt{2}}\left(
\left|\uparrow\downarrow\uparrow\cdots\right\rangle+
\left|\downarrow\uparrow\downarrow\cdots\right\rangle
\right). 
\end{equation}
The Majumdar-Ghosh state (dimer state)  is defined as 
\begin{equation}
|MG\rangle\equiv\frac{1}{2^{\frac{L}{4}}}
(\left|\uparrow\downarrow\right\rangle-
\left|\downarrow\uparrow\right\rangle)
(\left|\uparrow\downarrow\right\rangle-
\left|\downarrow\uparrow\right\rangle)\cdots. 
\end{equation}
%

\subsection{GHD solution of the expansion quench} 
\label{sec-ghd-eq}

It is interesting to discuss the solution of the expansion quench in the XXZ 
chain by using the framework of the Generalized Hydrodynamics (GHD) 
(see section~\ref{sec-ghd}). Since part $A$ of the system 
is prepared in the vacuum, a major simplification is  that 
\begin{equation}
\label{ss}
\vartheta_{+\infty,n}(\lambda)=0,\,\forall\lambda. 
\end{equation}
This has striking consequences in the continuity equation~\eqref{cont}. 
First, we restrict ourselves to the case $\zeta=0$, which, as already stressed, 
is the relevant ray to describe the interface between $A$ and $B$ and the 
steady-state entropies. 
By using the general solution~\eqref{ghd-sol} of~\eqref{cont}, Eq.~\eqref{ss} 
implies that $\vartheta_n$ is written as 
\begin{equation}
\label{ff}
\vartheta_{n}(\lambda)=\theta_H(v_{n})\vartheta_{-\infty,n}, 
\end{equation}
where we omit the subscript $\zeta$, since we are focusing on $\zeta=0$. 
Clearly, from~\eqref{ff}, one has that $\vartheta_{n}
(\lambda)$ is non-zero only for $\lambda$ such that $v_{n}
(\lambda)>0$. 
An important consequence of~\eqref{ff} is that 
$\vartheta_{n}(\lambda)=0$ for $\lambda$ such that $v_n(\lambda)<0$. 
The physical interpretation is that quasiparticles with these 
rapidities that are originated in $B$ do not reach the interface 
with $A$, because their group velocity change sign during 
the dynamics. It is useful to express Eq.~\eqref{ff} in terms of the dressed 
energies $e_n'$. By using the definition~\eqref{group-v}, 
and the fact that $\rho_n^{\scriptscriptstyle(t)}(\lambda)\ge0$, 
one has 
\begin{equation}
\label{cc}
v_n(\lambda)>0\Leftrightarrow e'_n(\lambda)>0, 
\end{equation}
which allows one to replace $\theta_H(v_n)$ with $\theta_H(e'_n)$ 
in~\eqref{ff}. An important consequence is that the integral 
equation for the dressed energy $e'_n$ (cf.~\eqref{v-int}) is 
decoupled from the continuity equation~\eqref{cont}, 
and it becomes 
\begin{multline}
\label{v-int-g}
0=\epsilon_n'-e_n'\\-\frac{1}{2\pi}
\sum_m\int d\mu e'_m(\mu)a_{nm}(\mu-\lambda)
\vartheta_{-\infty,m}(\mu)\theta_H(e'_m(\mu)).
\end{multline}
Here $\epsilon_n'$ is the derivative of the bare energy of the quasiparticles 
and $\vartheta_{-\infty,n}$ is the macrostate describing local equilibrium  
in the limit $\zeta\to-\infty$.

\subsection{Absence of bound-state transport after the expansion} 
\label{sec-ghd-eq1}

An intriguing feature of the expansion quench in the XXZ chain 
is that, for large enough $\Delta$, there is no bound-state transport 
between $A$ and $B$ (see Figure~\ref{cartoon}). This is a genuine effect 
of the interactions, which renormalize (``dress'') the group velocities of the 
system quasiparticles.  Due to this dressing, the 
group velocities $v_{n}$ of the bound states with $n>1$ that are 
created in $B$ change sign before reaching the boundary with $A$. 
Formally, the reason for this behavior is that for the XXZ chain 
in the large $\Delta$ limit one has 
\begin{equation}
\label{surp}
e'_n<0,\,\forall\lambda\quad\textrm{for}\,\,n>1.
\end{equation}
In the following, we first illustrate the mechanism by which 
Eq.~\eqref{surp} implies the absence of bound-state transport at 
large $\Delta$, and then we discuss its regime of validity upon 
lowering $\Delta$. 
Eq.~\eqref{surp} implies that in the system of integral equations~\eqref{v-int-g} 
the equation for $n=1$ is decoupled from the rest, and it is given as 
\begin{multline}
\label{e1}
\epsilon_1'=e_1'\\+\frac{1}{2\pi}\int 
d\mu e'_1(\mu)a_{11}(\mu-\lambda)\vartheta_{-\infty,1}(\mu)
\theta_H(e'_1(\mu)).
\end{multline}
The derivative $e_n'$ of the dressed energy density for bound states 
with $n>1$ is obtained by substituting the solution $e_1'$ 
of~\eqref{e1} in~\eqref{v-int-g}. This gives 
\begin{multline}
\label{cond}
e_n'=\epsilon'_n\\-\frac{1}{2\pi}\int d\mu e'_1(\mu)
a_{n1}(\mu-\lambda)\vartheta_{-\infty,1}(\mu)\theta_H(e'_1(\mu)).
\end{multline}
Eq.~\eqref{cond} has to be used to check~\eqref{surp} self-consistently 
(see below). 

The decoupling of the equation for $\eta_1$ in~\eqref{e1} is reflected 
in a similar behavior for the particle density $\rho_1$. From 
Eq.~\eqref{tba-eq}, it is clear that Eq.~\eqref{surp} and~\eqref{ff} imply 
that $\eta_n\to\infty$ and that $\rho_n\to0$ for $n>1$. On the other hand, 
for $n=1$, one has that $\rho_1$ is non-zero only for $\lambda$ such 
that $e_1'>0$. Specifically, from~\eqref{tba-eq} the finite part of 
$\rho_1$ is obtained by solving the integral equation 
\begin{multline}
\label{rho1}
\rho_1(\vartheta_{-\infty,1})^{-1}-\frac{a_1}{2\pi}
=\\-\frac{1}{2\pi}\int d\mu a_{11}(\lambda-\mu)\rho_1(\mu)\theta_{H}(e'_1(\mu)). 
\end{multline}
The fact that $\rho_n$ is identically zero for $n>1$ implies that 
the bound states with $n>1$ do not affect the local equilibrium 
properties at the interface between $A$ and $B$. 
Equivalently, the information about the bound 
states that are created in $B$ does not arrive at the interface with 
$A$. 

We now discuss the validity of this result as a function of $\Delta$. 
The strategy is to check~\eqref{surp} by substituting the solution of~\eqref{e1} 
in~\eqref{cond}. This gives the condition 
\begin{equation}
\label{cond1}
\epsilon'_n-\frac{1}{2\pi}\int d\mu e'_1(\mu)a_{n1}(\mu-\lambda)
\vartheta_{-\infty,1}(\mu)\theta_H(e'_1(\mu))<0. 
\end{equation}
We now show that for the expansion quench the condition~\eqref{cond1} is 
satisfied in the XXZ chain with large enough $\Delta$. First, one can 
verify that the leading order of the bare energy $\epsilon'_n$ for 
large $\Delta$ is given as~\cite{taka-book} 
\begin{equation}
\label{exp}
\epsilon_n'=2z^{n-1}\sin(2\lambda)+{\mathcal O}(z^{n+1}),\quad\textrm{with}\,
z\equiv e^{-\eta}. 
\end{equation}
Here $\eta\equiv\textrm{arccosh}(\Delta)$. 
We now assume that $e_1'={\mathcal O}(1)>0$ for 
some rapidities $\lambda$. We also assume that $\vartheta_{-\infty,1}>0$ 
and $\vartheta_{-\infty,1}={\mathcal O}(1)$. These conditions are 
verified for all the quenches considered in this work. It 
is also natural to expect that they hold true for a larger class 
of quenches. 
Now, using that in the large $\Delta$ limit, $a_{n1}={\mathcal O}(1)$ and $a_{n1}>0$, 
one has that the second term in~\eqref{cond1} is ${\mathcal O}(1)$ and negative, 
whereas the first one  can be made arbitrarily small upon increasing $\Delta$ 
(cf.~\eqref{exp}). This allows us to conclude that Eq.~\eqref{surp} holds true 
for large enough $\Delta$. Upon lowering $\Delta$, at a certain value $\Delta_n^*$ 
the $n$-particle bound states start to be transmitted between the two chains. 
These ``critical'' $\Delta^*_n$ depend on the initial state that is released. 
Furthermore, we observe that, at least for the first few values of $n$, 
$\Delta^*_n<\Delta_m^*$ for $n>m$, i.e., larger bound states start to be transmitted 
at smaller $\Delta$s. Finally, for the initial 
states that we consider (N\'eel state and Majumdar-Ghosh state) we notice that 
$\Delta_1^*<1.5$.

\subsection{Expansion quench in the large $\Delta$ limit: Exact results} 
\label{sec-ghd-eq2}

Before discussing the steady-state entropies, it is useful to investigate 
the large $\Delta$ expansion of the macrostate describing local and quasilocal 
observables. Here we focus on the expansion of the N\'eel state and the 
Majumdar-Ghosh state. 

\subsubsection{Expanding the N\'eel state}
\label{sec-lD-neel}

By using~\eqref{ff} and the fact that for the N\'eel state 
in the limit $\Delta\to\infty$ one has that~\cite{wdbf-14} 
$\vartheta_{-\infty,1}\to 1$, one obtains that $\vartheta_1$ 
becomes a step function. Precisely, one has 
\begin{equation}
\label{surp1}
\vartheta_1=\theta_H(e'_1(\lambda)). 
\end{equation}
Equation~\eqref{exp} and the fact that $a_{11}=2$ at the leading 
order in $1/\Delta$ imply that the dressed energy $e_1'$ is 
obtained by solving the integral equation  (cf.~\eqref{v-int-g})
\begin{equation}
\label{eq1}
e_1'=2\sin(2\lambda)-\frac{1}{\pi}\int d\mu e'_1(\mu)
\theta_H(e'_1(\mu)). 
\end{equation}
Eq.~\eqref{eq1} implies that 
\begin{equation}
\label{ans}
e_1'=2\sin(2\lambda)+\gamma,
\end{equation}
with $\gamma$ to be determined. After substituting~\eqref{ans} 
in~\eqref{eq1} one obtains an equation for $\gamma$ as 
\begin{multline}
\label{int}
2\sin(2\lambda)+\gamma=2\sin(2\lambda)\\-\frac{1}{\pi}
\int_{\lambda_-}^{\lambda_+}
d\mu (2\sin(2\lambda)+\gamma), 
\end{multline}
where we defined 
\begin{align}
\label{gg1}
& \lambda_-=-\frac{1}{2}\arcsin\Big(\frac{\gamma}{2}\Big)\\
\label{gg2}
& \lambda_+ =\frac{\pi}{2}+\frac{1}{2}\arcsin\Big(\frac{\gamma}{2}\Big). 
\end{align}
By performing the integral in~\eqref{int}, one obtains that $\gamma$ 
is the solution of 
\begin{equation}
\label{nonl1}
\frac{1}{\pi}\Big[\sqrt{4-\gamma^2}+\gamma\arccos\Big(\frac{\gamma}{2}\Big)\Big]+\gamma=0.
\end{equation}
Eq.~\eqref{nonl1} cannot be solved analytically. 
From~\eqref{nonl1} one obtains numerically $\gamma\approx -0.3989$. 

We now determine the particle density $\rho_1$. First, by using the TBA 
equation~\eqref{rho1} together with~\eqref{surp1}, one has that $\rho_1$ is zero 
for $\lambda\notin[\lambda_-,\lambda_+]$. For $\lambda\in[\lambda_-,\lambda_+]$, 
$\rho_1$ is obtained by solving 
\begin{equation}
\label{rho-eq}
\rho_1-\frac{1}{\pi}+\frac{1}{\pi}\int_{\lambda_-}^{\lambda_+}d\mu\rho_1=0, 
\end{equation}
where we used the TBA equation~\eqref{rho1}, and that at the leading order in 
$1/\Delta$ one has $a_1=2$. From~\eqref{rho-eq}, one has that $\rho_1$ 
exhibits a Fermi-sea structure, and it is given as 
%
%
%
\begin{equation}
\label{rho-n-d}
\rho_1=\left\{
\begin{array}{cc}
[\pi+\lambda_+-\lambda_-]^{-1} & \textrm{if}\, \lambda\in[\lambda_-,
\lambda_+]\\\\
0 & \textrm{otherwise}
\end{array}\right. 
\end{equation}
%
%
%
On the other hand, the hole density $\rho_1^{\scriptscriptstyle(h)}$ 
is nonzero only for $\lambda\notin[\lambda_-,\lambda_+]$, where 
$\rho_1^{\scriptscriptstyle(h)}=[\pi+\lambda_+-\lambda_-]^{-1}$. 
The fact that the hole and the particle density have complementary 
support implies that in the limit $\Delta\to\infty$ the 
Yang-Yang entropy is vanishing. 

\subsubsection{Expanding the Majumdar-Ghosh state}
\label{sec-lD-dimer}

A similar expansion can be obtained for the expansion of the 
Majumdar-Ghosh state. In contrast with the N\'eel state 
(see section~\ref{sec-lD-neel}), at the leading order 
in $1/\Delta$ one now has $\vartheta_{-\infty,1}=\cos^2(\lambda)$. 
The leading order in $1/\Delta$ of the dressed 
energy $e'_1$ (cf.~\eqref{e1}) is now obtained by solving 
\begin{equation}
\label{eq3}
e'_1=2\sin(2\lambda)-\frac{1}{\pi}\int d\mu e_1'(\mu)\cos^2(\mu)\theta_H(e_1'(\mu)). 
\end{equation}
Eq.~\eqref{eq3}  implies that 
\begin{equation}
e_1'=2\sin(2\lambda)+\gamma',
\end{equation}
where $\gamma'$ is obtained by solving 
\begin{equation}
\label{nonl2}
\frac{1}{2\pi}\Big[\sqrt{4-\gamma'^2}+\gamma'\arccos
\Big(\frac{\gamma'}{2}\Big)\Big]+\gamma'=0.
\end{equation}
Eq.~\eqref{nonl2} gives $\gamma'\approx -0.2487$. 
Eq.~\eqref{nonl2} is the same as~\eqref{nonl1}, apart from a 
factor $1/2$ in the first term. The TBA equation~\eqref{rho1} for 
the particle density $\rho_1$ is 
\begin{equation}
\label{cond2}
\frac{\rho_1}{\cos^2(\lambda)}-\frac{1}{\pi}+\frac{1}{\pi}
\int d\mu \rho_1(\mu)\theta_H(e_1'(\mu))=0. 
\end{equation}
This implies that $\rho_1=\gamma''\cos^2(\lambda)$. 
From~\eqref{cond2} the constant $\gamma''$ is obtained as   
%
%
%
\begin{equation}
\gamma''=\Big[\frac{5}{4}\pi-\frac{1}{2}\arcsin\Big(\frac{\gamma'
}{2}\Big)\Big]^{-1}, 
\end{equation}
where $\gamma'$ is the solution of~\eqref{nonl2}. 
Finally, the result for $\rho_1$ is written as 
%
%
%
\begin{equation}
\rho_1=\left\{
\begin{array}{cc}	
\cos^2(\lambda)\big[\frac{7}{4}\pi-\lambda_++\lambda_-\big]^{-1} & 
\textrm{if}\, \lambda\in[\lambda'_-,\lambda'_+]\\\\
0 & \textrm{otherwise}
\end{array}
\right.
\end{equation}
where we defined
\begin{align}
\label{ggg1}
	& \lambda'_-=-\frac{1}{2}\arcsin\Big(\frac{\gamma'}{2}\Big)\\
\label{ggg1x}
	& \lambda'_+ =\frac{\pi}{2}+\frac{1}{2}\arcsin\Big(\frac{\gamma'}{2}\Big). 
\end{align}
Notice that $\lambda_\pm'$ coincide with $\lambda_\pm$ (cf.~\eqref{gg1}\eqref{gg2}) 
after the substitution $\gamma'\to\gamma$. Clearly, in contrast with the 
N\'eel state (see section~\ref{sec-lD-neel}), $\rho_1$ is not a step function.

\subsection{R\'enyi entropies after the expansion quench}
\label{sec-eq-renyi}

We now calculate the steady-state entropies after 
the expansion quench. We restrict ourselves to $\Delta>\Delta_2^*$ 
(see section~\ref{sec-ghd-eq1}), which allows us to neglect the 
contribution of the multispin bound states with $n>1$. 

As described in section~\ref{sec-ghd-main}, the first step is to determine 
the driving functions $g_n$ (see Eq.~\eqref{s-eta} and 
Eq.~\eqref{m-eta}). Here $g_n$ is obtained by first solving~\eqref{v-int-g} 
for $e_n'$ and then by substituting in 
\begin{multline}
\label{reff}
g_n=\ln([\vartheta_{-\infty,n}\theta_H(e'_n)]^{-1}-1)\\
+\sum_m\int d\mu a_{nm}(\mu-\lambda)
\ln(1-\vartheta_{-\infty,m}(\mu)\theta_H(e'_m)), 
\end{multline}
where we used that for the expansion quench $\vartheta_n=\theta_H(e_n')
\vartheta_{-\infty, n}$. 

By using that $e_n'<0$ for $n>1$, 
one obtains that $g_n$ is divergent for any $\lambda$ for $n>1$, 
whereas $g_1$ diverges for $\lambda$ such that $e'_1<0$, and it 
is finite otherwise. 
To derive the R\'enyi entropies one has now to solve the TBA 
equations with the modified driving $\alpha g_n$. As a consequence of the 
results outlined above, some simplifications occur. Let us consider 
the integral equations~\eqref{m-eta} for $\eta_n^{\scriptscriptstyle
(\alpha)}$
\begin{multline}
\label{a-driv}
\ln(\eta_n^{(\alpha)})=\alpha g_n\\
+\sum\limits_{m=1}^\infty \int d\mu a_{nm}(\mu-\lambda) 
\ln(1+1/\eta^{(\alpha)}_m(\mu)). 
\end{multline}
First, in Eq.~\eqref{a-driv}, the integral in the right hand side is 
nonzero only if $\eta^{\scriptscriptstyle(\alpha)}_m$ is finite. 
Thus, from~\eqref{a-driv}, one has that $\eta_n=\alpha g_n$ 
are divergent for $m>1$ because $g_n$ are divergent. On the other 
hand, $\eta_1$ is finite for $\lambda$ such that 
$g_1$ is finite, i.e., if $e'_1(\lambda)>0$ 
(cf.~\eqref{reff}), 
and it is divergent otherwise. This implies that the equation for $\eta^{\scriptscriptstyle(\alpha)}_1$ 
is decoupled from the rest and it is given as 
\begin{multline}
\label{etaa-eq}
\ln(\eta_1^{(\alpha)})=\alpha g_1\theta_H(e_1')\\
+\int d\mu a_{11}(\mu-\lambda)\ln(1+1/\eta^{(\alpha)}_1(\mu)
\theta_H(e_1')). 
\end{multline}
Here $e_1'$ is obtained by solving~\eqref{e1}. 
We now discuss the macrostate densities $\rho^{\scriptscriptstyle(
\alpha)}_n$.  The TBA equations~\eqref{tba-eq} become  
\begin{equation}
2\pi\rho_n^{(\alpha)}(1+\eta^{(\alpha)}_n)=a_n-\sum_m\int d\mu a_{nm}(\mu-\lambda)
\rho^{(\alpha)}_m(\mu). 
\end{equation}
Clearly, $\rho^{\scriptscriptstyle(\alpha)}_n$ is zero for $n>1$, 
reflecting the divergent behavior of $\eta_n^{\scriptscriptstyle(\alpha)}$. 
Also, $\rho_1^{\scriptscriptstyle(\alpha)}$ is non-zero only for $\lambda$ 
such that $\eta_1^{\scriptscriptstyle(\alpha)}$ is finite. To obtain 
the finite part of $\rho_1^{\scriptscriptstyle(\alpha)}$ one has to 
solve the integral equation 
\begin{equation}
2\pi\rho_1^{(\alpha)}(1+\eta^{(\alpha)}_1)=a_1-\int d\mu a_{11}(\mu-\lambda)
\rho^{(\alpha)}_1(\mu)\theta_H(e'_1). 
\end{equation}
It is interesting to observe that for any finite $\alpha$ the support of 
$\rho_1^{\scriptscriptstyle(\alpha)}$ does not depend on $\alpha$, but it 
is determined by the condition $e_1'>0$. 


It is now worth making some remarks on the treatment of the single 
copy entanglement for $\alpha\to\infty$.  
Similar to finite $\alpha$, one has to determine the TBA density 
$\eta^{\scriptscriptstyle(\infty)}_1$. The ansatz~\eqref{ansatz} 
becomes 
\begin{equation}
\eta_1^{(\alpha)}=\exp(\alpha \gamma_1), 
\end{equation}
%
%
\begin{figure}[t]
\includegraphics*[width=1\linewidth]{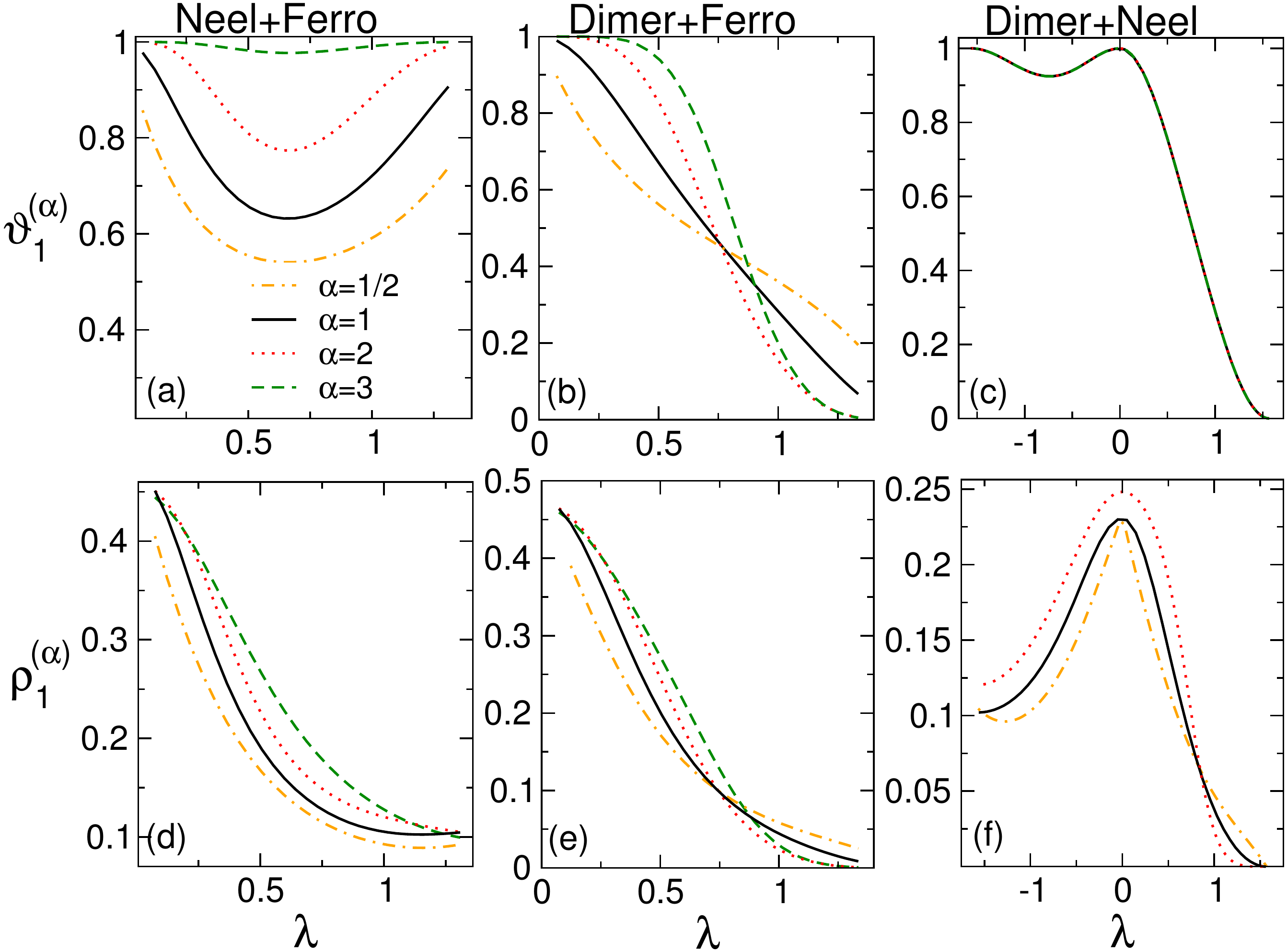}
\caption{ R\'enyi entropies after a quench from a piecewise homogeneous 
 initial state in the XXZ chain. 
 Filling function $\vartheta^{\scriptscriptstyle(\alpha)}_1$ and 
 particle density $\rho_1^{\scriptscriptstyle(\alpha)}$ (panel (a-c) and 
 (d-e), respectively) describing steady-state R\'enyi entropies. On the $x$-axis $\lambda$ is the 
 quasiparticles rapidity. Panels in different columns correspond to quenches 
 from different initial states. 
 All the results are for $\Delta=2$. 
 For the expansion quenches (panels (a,b) and (d,e)) 
 only the $n=1$ string densities are nonzero. For the 
 quench from the state $|N\otimes MG\rangle$ higher strings contribute 
 (not reported in the figure). 
}
\label{fig1a}
\end{figure}
%
where the function $\gamma_1$ has to be determined. 
One has that in the limit $\alpha\to\infty$, Eq.~\eqref{integ1} 
becomes 
\begin{equation}
\label{g1}
\gamma_1=g_1-\int d\mu a_{11}(\mu-\lambda)\gamma^+_1(\mu), 
\end{equation}
where $\gamma_1^+$ is defined in~\eqref{constr}. 
The particle density $\rho_1^{\scriptscriptstyle(\infty)}$ 
(cf.~\eqref{rho-in}) becomes 
\begin{equation}
\label{ainf}
2\pi\rho_1^{(\infty)}=a_1-\int d\mu a_{11}(\mu-\lambda)\rho_1^{(\infty)}(\mu)
\theta_H(-\gamma_1). 
\end{equation}
%
The nonzero part of $\rho_1^{\scriptscriptstyle(h,\infty)}$ is obtained by
using~\eqref{rhoh-in}. 
%
%


\subsection{Numerical TBA results}
\label{sec-xxz-num}

We now provide exact numerical results for the steady-state 
entropies after the quench from a piecewise homogeneous initial state in 
the XXZ chain. 

\subsubsection{Macrostate densities and ``critical'' anisotropies}
\label{sec-xxz-den}

Numerical results for the macrostate densities $\vartheta_n^{
\scriptscriptstyle(\alpha)}$ and 
$\rho_n^{\scriptscriptstyle(\alpha)}$ are shown in Figure~\ref{fig1a}, 
for the expansion quenches of the N\'eel state and of 
the Majumdar-Ghosh state  
(first and second column, respectively), and for the quench from 
$|N\otimes MG\rangle$ (third column in Figure~\ref{fig1a}). 
For the expansion quenches we restrict ourselves to $\Delta>\Delta_2^*$ 
(see section~\eqref{sec-ghd-eq1} for 
its definition). As it has been already discussed, this implies that  only 
the densities $\vartheta_1^{\scriptscriptstyle(\alpha)}$ and 
$\rho_1^{\scriptscriptstyle(\alpha)}$ are nonzero. For 
$|N\otimes MG\rangle$ we show only results for $n=1$, 
although all the densities with $n>1$ are nonzero. 
Also, one should observe that for the expansion quenches 
the densities are nonzero only in 
a subset of $[-\pi/2,\pi/2]$, whereas for $|N\otimes MG\rangle$ they 
are nonzero in the full interval $[-\pi/2,\pi/2]$. 
Finally, as it is clear from panel (c), $\vartheta_1$ 
exhibits a quite weak dependence on $\alpha$. 

%
\begin{figure}[t]
\includegraphics*[width=0.99\linewidth]{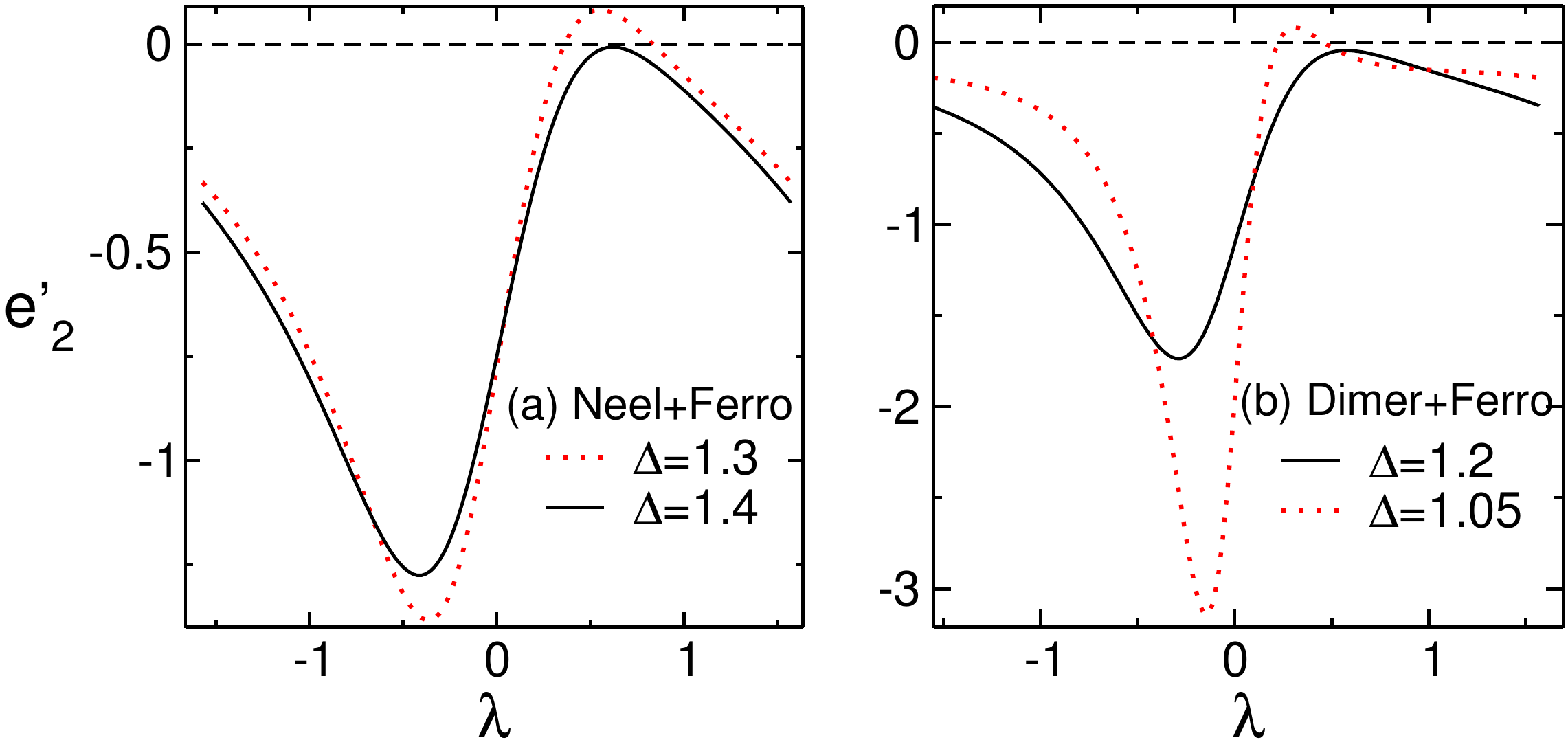}
\caption{``Critical'' anisotropies $\Delta_2^*$ for the two-particle 
 transport after a piecewise homogeneous quench in the XXZ chain. 
 The two panels are for 
 the expansion of the N\'eel and the the Majumdar-Ghosh states. 
 The figure shows the derivative of dressed energy $e'_2$ 
 (cf.~\eqref{v-int}) for $n=2$, 
 plotted as a function of $\lambda$. For $\Delta>\Delta_2^*$, 
 one has that $e_2'<0,\,\forall\lambda$, which ensures that 
 transport of two-particle bound states is inhibited. Note 
 that $\Delta_2^*\approx 1.4$ and $\Delta_2^*\approx1.15$ 
 in (a) and (b), respectively. 
}
\label{fig-trans}
\end{figure}
%

It is interesting to extract the ``critical'' anisotropies $\Delta_n^*$ for 
the expansion quenches. 
Here we focus on $\Delta_2^*$. 
As we discussed in section~\ref{sec-ghd-eq1}, for $\Delta>\Delta_2^*$ one 
has that $e'_2<0$, giving $\vartheta_2^{\scriptscriptstyle(\alpha)}=0$, 
which implies that transport of the two-particle bound states between 
$A$ and $B$ is absent. Formally, $\Delta^*_2$ is the point at which $e_2'$ 
becomes positive. 
This is discussed in Figure~\ref{fig-trans} showing $e_2'$ as 
a function of $\lambda$, for the expansion of the 
N\'eel state and the Majumdar-Ghosh state. 
For the N\'eel state (panel (a)), one has that for any $\lambda$, $e'_2<0$ for 
$\Delta\gtrsim 1.4$, whereas $e_2'>0$ for $\Delta=1.3$, which suggests 
that $1.3\gtrsim\Delta_2^*\lesssim1.4$. For the Majumdar-Ghosh state 
one has a much smaller values for $\Delta_2^*$. Indeed, from panel 
(b) it is clear that $\Delta_2^*<1.1$. 
The same analysis could be performed for the three-particle bound states, 
i.e., for $n=3$. One should expect 
a new ``critical'' anisotropy $\Delta_3^*\le\Delta^*_2$. For 
$\Delta<\Delta_3^*$ transport of the two-particle  and three-particle 
bound states is permitted. We numerically 
observed that $e_3\ll e_2$ for $\Delta<\Delta_2^*$, suggesting that 
there is an extended region in $\Delta$ where only transport of $2$-particle and 
$3$-particle bound states is allowed. It is natural to expect that a similar 
scenario occurs for larger bound states, i.e., that there is a ``cascade'' of 
transition points $\Delta_1^*>\Delta_2^*>\Delta_3^*>\Delta_4^*\dots$. This would 
lead to the intriguing scenario in which transport of larger and larger 
bound states is activated at smaller and smaller anisotropies. 

%
\begin{figure}[t]
\includegraphics*[width=.95\linewidth]{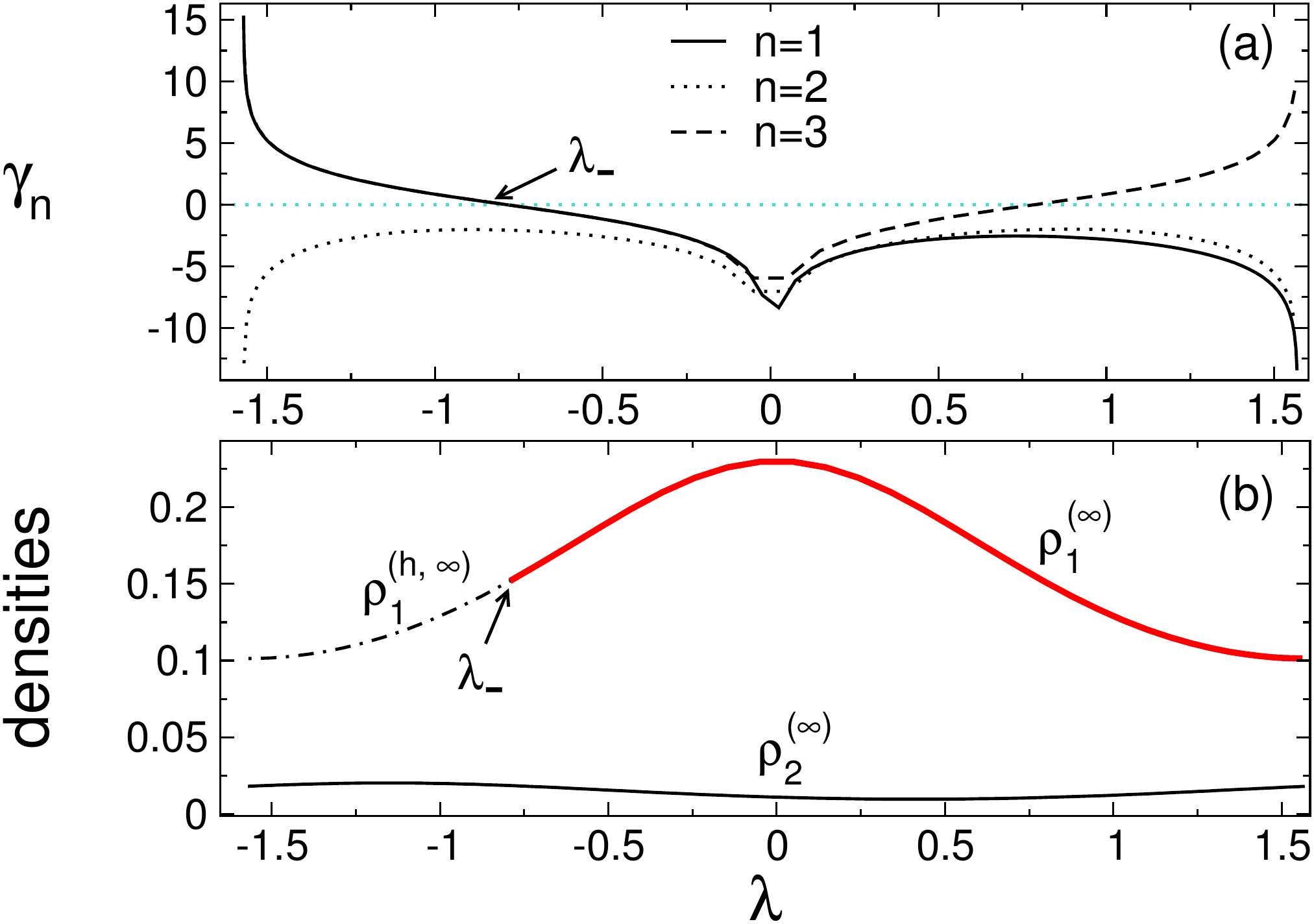}
\caption{Min entropy after a piecewise homogeneous quench 
 in the XXZ chain: Quench from the state $|N\otimes MG\rangle$, 
 for $\Delta=5$. Panel (a) shows $\gamma_n$ versus $\lambda$ 
 for $n\le3$. Note that $\gamma_2<0$ for any $\lambda$, in contrast 
 with $\gamma_1$ and $\gamma_3$. 
 Panel (b) shows $\rho^{\scriptscriptstyle(\infty)}_n$ 
 for $n=1,2$. Continuous and dashed-dotted lines denote 
 particle and hole densities, respectively. 
}
\label{scopy}
\end{figure}
%

%
\begin{figure*}[t]
\includegraphics*[width=0.93\linewidth]{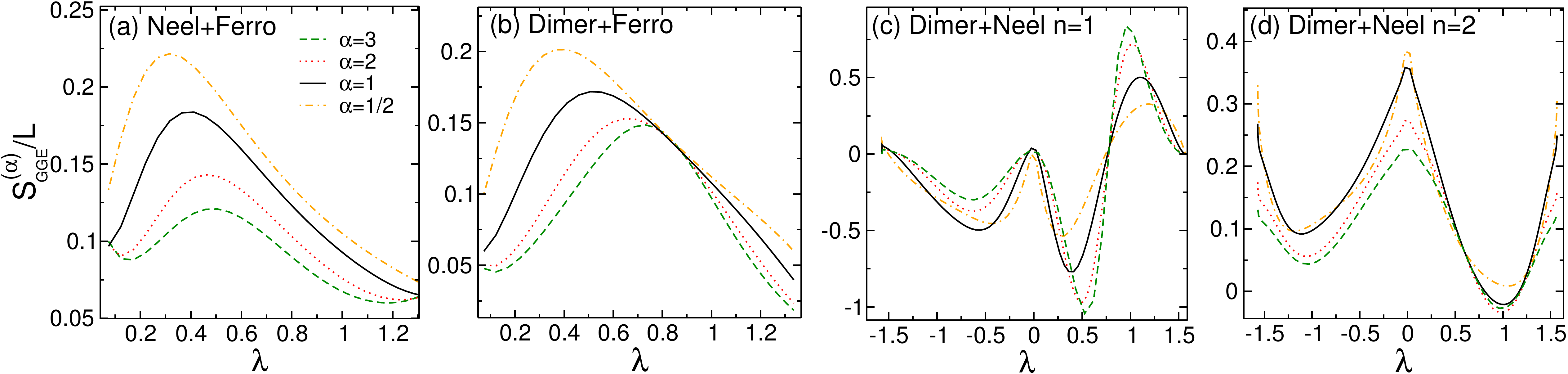}
\caption{ Steady-state R\'enyi entropies after the quench from a piecewise 
 homogenous initial state in the XXZ chain : 
 Quasiparticle contributions. The entropy density 
 $S_{\textrm{GGE}}^{\scriptscriptstyle (\alpha)}/L$ is plotted 
 versus $\lambda$, for several values of $\alpha$. 
 Panels (a) and (b) are for the expansion quench of the N\'eel state 
 and the Majumdar-Ghosh state, respectively. Data are for $\Delta=2$. 
 The entropy density is positive for all $\lambda$. Panels (c) 
 and (d) show results for the quench from $|N\otimes MG\rangle$ 
 for $\Delta=5$. Only results for $n=1$ and $n=2$ are shown. 
}
\label{fig3}
\end{figure*}
%

Finally, it is useful to consider the limit $\alpha\to\infty$. 
We focus on the quench from $|N\otimes MG\rangle$. 
Figure~\ref{scopy} shows 
the TBA densities $\gamma_n$ (see~\eqref{ansatz} for 
their definitions), $\rho^{\scriptscriptstyle(\infty)}_n$ 
and $\rho^{\scriptscriptstyle(h,\infty)}$ (panels (a) and (b) 
plotted as function of $\lambda$. 
Results are for  $\Delta=5$. In panel (a) we only 
show results for $\gamma_n$ with $n\le 3$. Crucially, one has 
that $\gamma_1<0$ for $\lambda>\lambda_-$, with $\lambda_-
\approx-0.75$. Similarly, $\gamma_3$ is negative in a subinterval 
of $[-\pi/2,\pi/2]$. On the other hand, one has that $\gamma_2<0$ 
for any $\lambda\in[-\pi/2,\pi/2]$. This is reflected in the 
behavior of $\rho_n^{\scriptscriptstyle(\infty)}$ (panel (b)). 

The continuous lines in panel (b) are the particle densities 
$\rho^{\scriptscriptstyle(\infty)}_1$ and $\rho^{\scriptscriptstyle
(\infty)}_2$. The support of 
$\rho_2^{\scriptscriptstyle(\infty)}$ is the full interval
$[-\pi/2,\pi/2]$, whereas $\rho^{\scriptscriptstyle(\infty)}_1$ is 
non-zero only for $\lambda>
\lambda_-$, with $\lambda_-$ the same as in panel (a). The complementary 
part of the curve (dashed-dotted line) in the panel is the hole 
density $\rho_1^{\scriptscriptstyle(h,\infty)}$. 

\subsubsection{R\'enyi entropies}
\label{sec-xxz-ent}

The densities reported in Figure~\ref{fig1a} are used to calculate the steady-state 
entropies by using~\eqref{main}. 
Figure~\ref{fig3} shows the contributions of the individual quasiparticles to the 
entropies, plotted versus $\lambda$. We report results for 
the expansion of the N\'eel state and the Majumdar-Ghosh 
state (in (a) and (b)). Data for the quench from $|N\otimes MG\rangle$ and for  
$n=1$ and $n=2$ are shown in (c) and (d). 
Notice, however, that for $|N\otimes MG\rangle$, bound states 
with $n>1$ contribute to the R\'enyi entropies, although we do not 
report them in the figure. 
Data in panel (a) and (b) are for the XXZ chain 
with $\Delta=2$, while in (c) and (d) we consider $\Delta=5$. 
While the entropy density for the expansion  
quenches is positive for all values of $\lambda$ (see panels (a) and (b)), this is 
not the case for $|N\otimes MG\rangle$. 
The same behavior, i.e., that the density of R\'enyi entropies 
is not positive, was  observed in homogeneous quenches, and it is 
the main obstacle when applying the quasiparticle picture to describe 
the full-time dynamics of the entropies~\cite{alba-2017,alba-2017-a}. 
%
\begin{figure}[b]
\includegraphics*[width=1\linewidth]{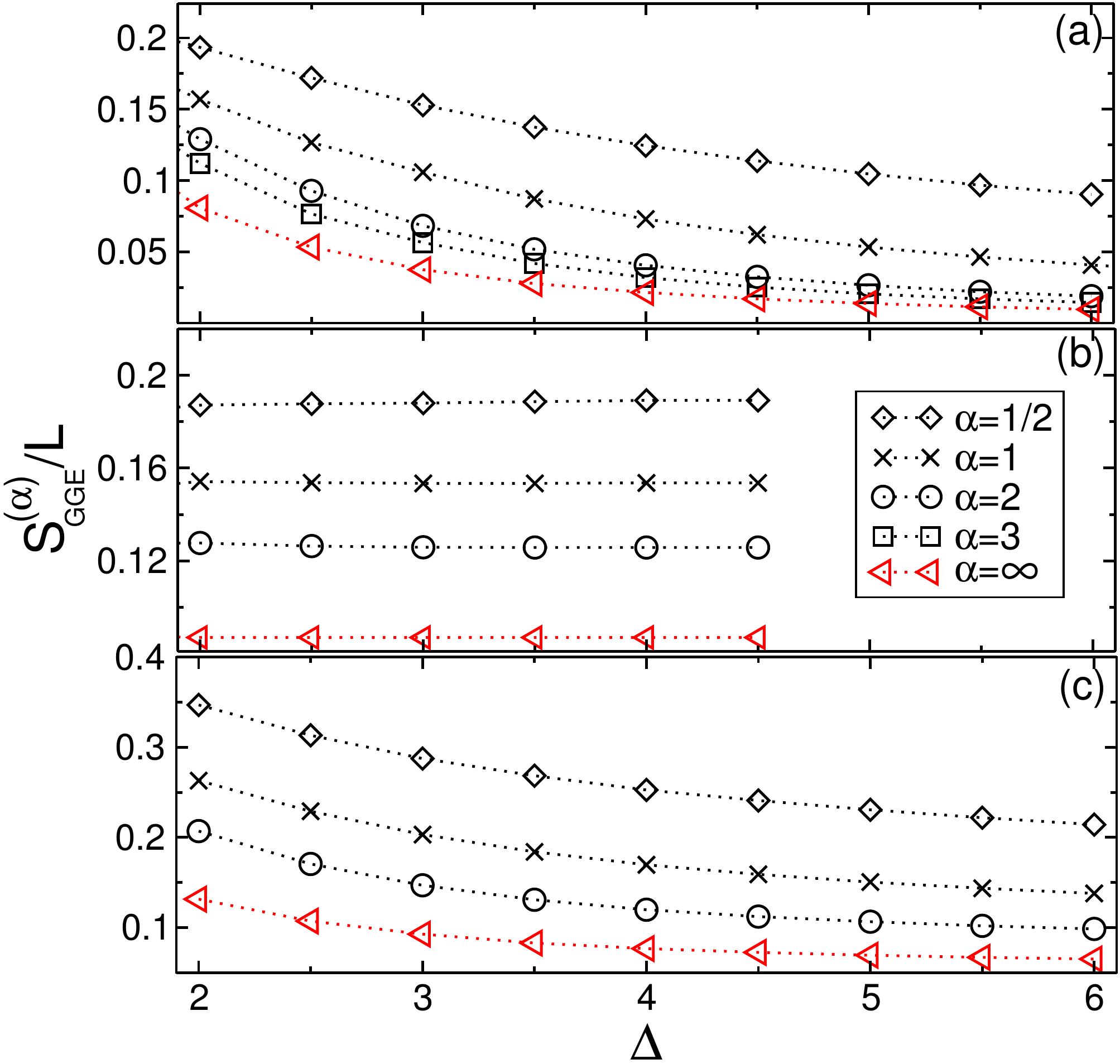}
\caption{ Steady-state R\'enyi entropies after a quench from a 
 piecewise homogeneous initial state in the XXZ chain: GHD results 
 for different initial states and R\'enyi index $\alpha$. 
 The entropy density $S_{\textrm{GGE}}^{\scriptscriptstyle (\alpha)}/L$, 
 is plotted against $\Delta$. In (a) and (b) the triangle 
 correspond to the min entropy ($\alpha\to\infty$) 
}
\label{fig7a}
\end{figure}
%
The GHD prediction for the steady-state entropis is obtained by integrating 
the results in Figure~\ref{fig3} and by summing over the quasiparticle 
families. 
The results are reported in Figure~\ref{fig7a} as a function of $\Delta$, 
and for different initial states. 
For all the quenches one has that $S^{\scriptscriptstyle(\alpha)}<
S^{\scriptscriptstyle(\alpha')}$, for $\alpha'<\alpha$, as expected. 
For the expansion of the N\'eel state, all the R\'enyi entropies vanish in 
the large $\Delta$ limit. This reflects that the N\'eel state is the ground state of 
the XXZ chain in that limit, similar to the homogeneous case~\cite{alba-2017-a}. 
For the expansion of the Majumdar-Ghosh state the 
entropies exhibit a quite weak dependence on $\Delta$. 
Results for the quench from $|N\otimes MG\rangle$ are reported in 
panel (c). The entropy densities exhibit a decreasing trend upon 
increasing $\Delta$, although they do not vanish in the limit 
$\Delta\to\infty$.

\section{DMRG results for the XXZ chain}
\label{sec-numerics}

%
\begin{figure*}[t]
\includegraphics*[width=0.9\linewidth]{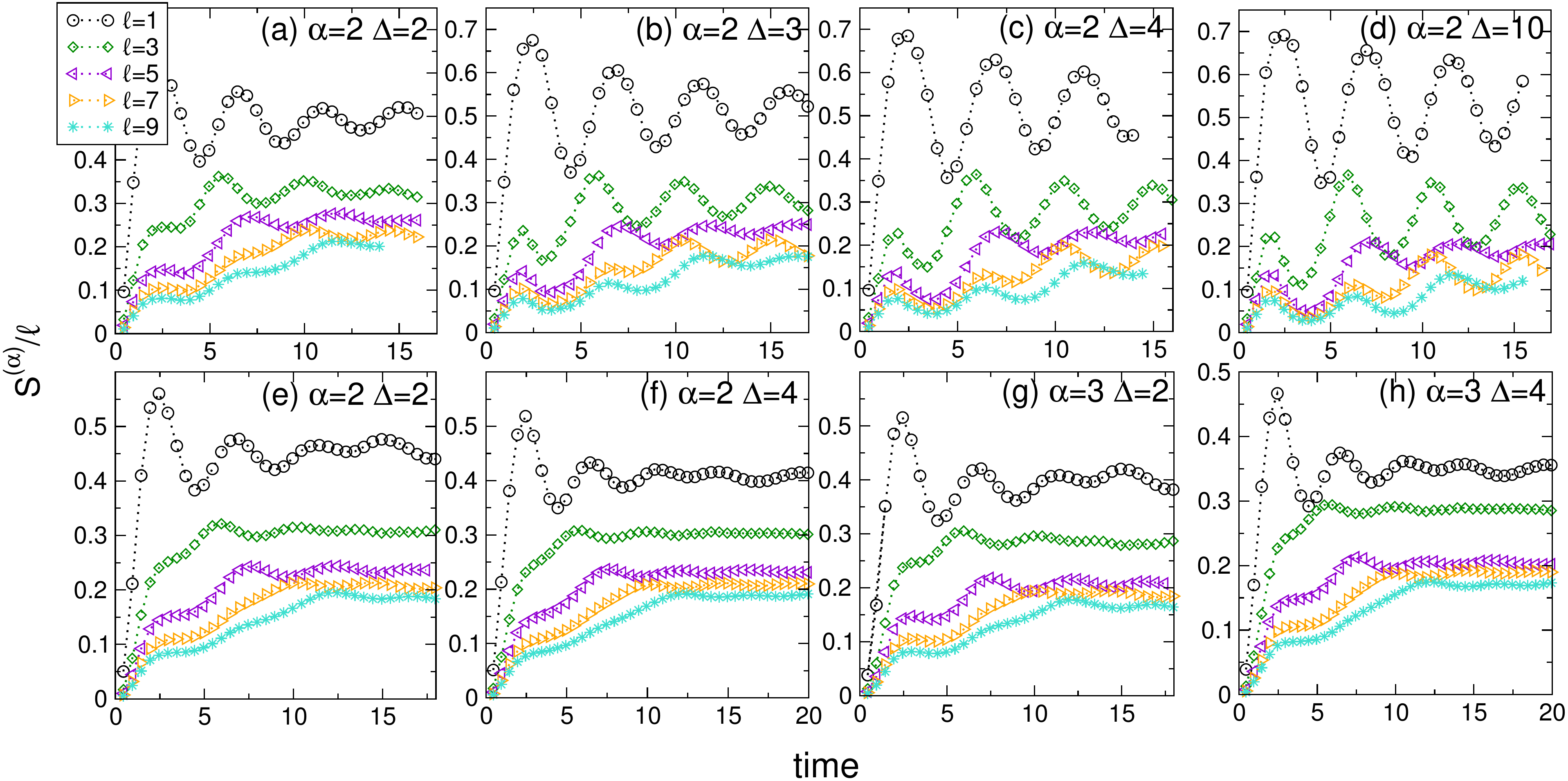}
\caption{ Steady-state R\'enyi entropy after the expansion quench of the 
 N\'eel and the Majumdar-Ghosh state (panels (a-d) and 
 (e-h), respectively) in the XXZ chain: tDMRG 
 results. The entropy density $S^{\scriptscriptstyle(\alpha)}/\ell$ is 
 plotted versus the time after the quench. For the expansion quench of 
 the N\'eel state (panels (a-d)) only $S^{\scriptscriptstyle(2)}$ is shown. 
 For the expansion quench of the Majumdar-Ghosh state panels (c,d) show results 
 for $S^{\scriptscriptstyle(2)}/\ell$, whereas (e,f) are for 
 $S^{\scriptscriptstyle(3)}/\ell$. 
 In all panels different symbols  are used for different values of the subsystem 
 size $\ell$. Only odd values of $\ell$ are reported. Notice the large oscillations 
 upon increasing $\Delta$ in panels (a-d). 
}
\label{fig8}
\end{figure*}
%

We now turn to compare the TBA results presented in section~\ref{sec-xxz} against 
tDMRG simulations~\cite{itensor}. We restrict ourselves to the expansion quenches of 
the N\'eel and the Majumdar-Ghosh states, as they are easier to simulate with 
tDMRG. We employ the framework of the Matrix Product States (MPS). All 
the considered initial states admit a simple MPS representation with rather small 
bond dimension $\chi$. The initial states are time-evolved by using a standard 
second order Trotter-Suzuki decomposition of the evolution operator $e^{-i Ht}$. 
The Trotter time discretization step is $\delta t=0.05$. At each step of 
the evolution the state looses its MPS form, which has to be restored by 
performing a singular value decomposition (SVD). To prevent the rapid growth 
of $\chi$, a truncated SVD is performed with maximum allowed bond dimension 
$\chi_{max}$. In our simulations we employed $\chi_{max}=80$. Importantly, 
here we are interested in calculating the R\'enyi entropies of a subsystem 
embedded in the chain (see Figure~\ref{cartoon}). In the standard MPS 
framework, the computational cost for calculating the entropy is $\propto\chi^6$ 
(see Ref.~\onlinecite{paola-2016} for the details), in contrast with the  
cost for calculating the half-chain entropy, which is only $\propto\chi^3$.  

Our tDMRG data are reported in Figure~\ref{fig8} for both the 
N\'eel state and of the Majumdar-Ghosh state. The Figure shows the 
entropy density $S^{\scriptscriptstyle(\alpha)}/\ell$ 
versus time. All the results are for the XXZ chain with $L=40$. 
Different curves correspond to different values of $\ell$. For clarity, 
we only show results for odd $\ell$. 
Panels (a-d) show data for the N\'eel state and 
for $\alpha=2$ and different values of 
$\Delta$. Panels (e-h) show the results for the 
Majumdar-Ghosh state. Specifically, panels (e-f) plot 
$S^{\scriptscriptstyle(2)}/\ell$ for $\Delta=2$ and $\Delta=4$, whereas 
panels (g-h) are for $S^{\scriptscriptstyle(3)}/\ell$ for the same values 
of $\Delta$. 
For the N\'eel state, the entropy density decreases upon increasing 
$\Delta$, as expected in the scaling limit (see Figure~\ref{fig7a}). Interestingly, 
sizeable oscillations with time are observed, whose amplitude increases 
with increasing $\Delta$. This is similar to the homogeneous 
quench from the N\'eel state~\cite{alba-2017-a}. 
On the other hand, for the Majumdar-Ghosh state, 
although oscillations with time are present, they decay 
quite rapidly with increasing $\ell$, 
and their amplitude does not increase upon increasing $\Delta$. 
%
\begin{figure*}[t]
\includegraphics*[width=0.93\linewidth]{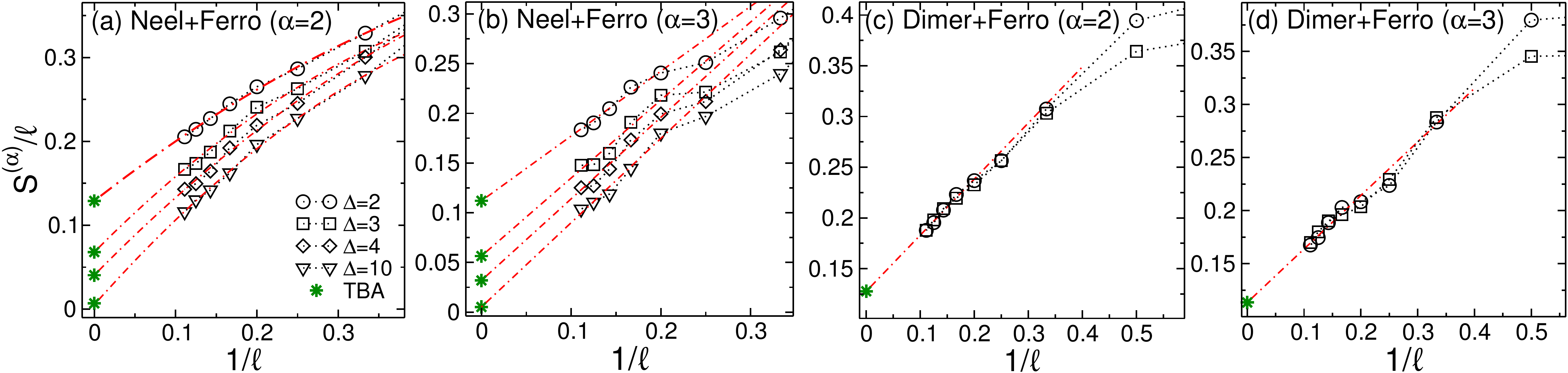}
\caption{ Steady-state R\'enyi entropies after a quench from a 
 piecewise homogeneous initial state in the XXZ chain: 
 Comparison between theory and tDMRG simulations. Panels (a) and (b) 
 are for the expansion of the N\'eel state for $\alpha=2$ 
 and $\alpha=3$, respectively. In all panels $S^{\scriptscriptstyle(\alpha)}/\ell$ 
 is plotted versus $1/\ell$, with $\ell$ the subsystem size. 
 The star symbols are the 
 TBA predictions in the scaling limit. The dashed-dotted lines are fits to 
 $S^{\scriptscriptstyle(\alpha)}/\ell=s^{\scriptscriptstyle(\alpha)}_\infty+
 a/\ell+b/\ell^2$, with $a,b$ fitting parameters and 
 $s_\infty^{\scriptscriptstyle(\alpha)}$ 
 fixed by the TBA result. Panels (c) and (d): Same as in (a) and (b) 
 for the expansion of the Majumdar-Ghosh state. Data are now 
 for $\Delta=5$. Notice in panels (c) and (d) the weak dependence 
 on $\Delta$. 
}
\label{fig3a}
\end{figure*}
%
The comparison between the Bethe ansatz results for the R\'enyi 
entropies is presented in Figure~\ref{fig3}. Panels (a)-(d) show 
$S^{\scriptscriptstyle (\alpha)}/\ell$ for $\alpha=2,3$. 
In all the panels the different symbols are tDMRG data for different 
values of $\Delta\in[2,10]$ and $\ell\le 10$. On the $x$-axis we 
show $1/\ell$. The raw tDMRG data are reported 
in Figure~\ref{fig8}. The results presented in Figure~\ref{fig3} 
are obtained by averaging the tDMRG results for $t>10$, to mitigate 
the effect of the oscillations with time. The star symbols are the TBA 
results in the scaling limit (see Figure~\ref{fig7a}). Clearly, due 
to the finite $\ell$, corrections are visible. To recover the 
scaling limit results, we perform a finite-size scaling analysis. 
The dashed-dotted lines in Figure~\ref{fig3} are fits to 
\begin{equation}
S^{(\alpha)}=\Big[s_\infty^{\scriptscriptstyle(\alpha)}+
\frac{a}{\ell}+\frac{b}{\ell^2}\Big]\ell, 
\end{equation}
where $s_\infty^{\scriptscriptstyle(\alpha)}$ is the TBA prediction and 
$a,b$ are fitting parameters. In all the panels the fits confirm that 
in the scaling limit the steady-state R\'enyi entropies 
are described by the TBA results.

\section{Conclusions}
\label{sec-concl}

We investigated the steady-state R\'enyi entropies after a quench 
from a piecewise homogeneous initial state in interacting integrable 
models. Our results were obtained by combining the TBA approach for the 
R\'enyi entropies developed in Ref.~\onlinecite{alba-2017} and the 
GHD treatment of the quench. We provided explicit 
results for quenches in the anisotropic Heisenberg XXZ chain from several 
initial states. We benchmarked our results against tDMRG simulations, 
finding always satisfactory agreement. 
We also investigated the steady-state entropies after 
the expansion quench, in which one of the two chains is prepared in the 
vacuum of the model excitations. Interestingly, we observed that for large 
enough anisotropy the transport of multi-spin bound states is not allowed. 
This is reflected in the steady-state entropies, which 
do not contain information about the bound states. 
On the other hand, we showed that there is a ``critical'' anisotropy 
below which bound-state transport is allowed. 

We now discuss some open problems and possible directions for future 
work. First, it is important to extend the quasiparticle picture 
to describe the full-time dynamics of the R\'enyi entropies after 
the quench. To this purpose, two main issues appear. First, 
similar to the von Neumann entropy, describing the full-time dynamics 
of the R\'enyi entropies requires determining the trajectories of the 
quasiparticles as function of time, which implies that rays with 
$\zeta\ne0$ have to be considered. This analysis has been 
performed in Ref.~\cite{to-write} for the von Neumann entropy. 
A more severe obstacle is that the TBA macrostate that describes the steady-state entropies is not 
the same as that describing (quasi)local observables and the von Neumann 
entropy, similar to what happens for homogeneous quenches~\cite{alba-2017-a}. 
This does not provide the correct framework for describing the entropy dyanamics, 
because the excitations that are responsible of the entanglement growth 
are the ones constructed around the steady state. To apply the quasiparticle 
description to the R\'enyi entropies one would first need to express the 
entropy densities as function of the TBA macrostate that describes the 
steady state. This, however, is still an open and difficult problem. 

It would be also important to extend our analysis to different initial 
states, such as the tilted N\'eel state and the tilted ferromagnet. 
For the latter case, it would be interesting to investigate the 
effect of the tilting on the bound-state transport between the two chains. 
Another interesting direction is to consider piecewise homogeneous quenches 
in continuum systems such as the Lieb-Liniger model, or in the Hubbard chain 
in the limit of strong interactions~\cite{BeTC17}. 

\section{Acknowledgements} 

I  acknowledge very fruitful discussions with
Benjamin Doyon. I am grateful to  Maurizio Fagotti and 
Bruno Bertini for several discussions on a related project. 
I thank Pasquale Calabrese for reading the manuscript and 
for several comments. 
This work was supported by the European
Union's  Horizon  2020  research  and  innovation  programme under the Marie
Sklodowoska-Curie grant agreement No 702612 OEMBS.

\end{document}